\documentclass[12pt]{article}\synctex=1 
\usepackage{amsfonts}
\usepackage{amssymb}
\usepackage{amsmath}
\usepackage{amsthm}
\usepackage{geometry}
\usepackage{commath}
\usepackage{booktabs}
\usepackage{lscape}
\usepackage[flushleft]{threeparttable}
\usepackage{bigstrut}
\usepackage{xr}

\usepackage[authoryear]{natbib}

\newcommand{\myref}[2]{\hyperref[#1]{#2}}
\numberwithin{equation}{section}

\newtheorem{assum}{Assumption}

\numberwithin{assum}{section}

\numberwithin{thm}{section}
\usepackage{latexsym}
\usepackage{graphicx} 
\usepackage{subfig}
\graphicspath{ {figures/} }
\usepackage{enumerate}
\usepackage{setspace}
\usepackage[toc,title,titletoc,header]{appendix}
\usepackage[bottom]{footmisc}   
\usepackage[pdfborder={0 0 0}]{hyperref}
\usepackage{float}
\usepackage{threeparttable}
\usepackage{lineno}
\setpagewiselinenumbers

\theoremstyle{definition}

\theoremstyle{remark}

\newcounter{assumptionA}
\def\theassumptionA{\arabic{assumptionA}}
\newcounter{theorema}

\def\thetheorem{\arabic{theorema}}
\newcounter{definitiona}

\def\thedefinition{\arabic{definitiona}}
\hypersetup{
  colorlinks=true,linkcolor=blue, citecolor=blue, filecolor=blue, 
  linkbordercolor={0 1 1}, citebordercolor={1 0 0},
  pdfcreator={\LaTeX\ with package \flqq hyperref\frqq},
  pdfsubject={},
}

\begin{document}

{
\title{A Practitioner's Guide to AI+ML in Portfolio Investing}
\author{\textsc{Mehmet Caner\thanks{
			North Carolina State University, Nelson Hall, Department of Economics, NC 27695, Magnus-AI-Turkiye. Email: mcaner@ncsu.edu. }}
	\and \textsc {Qingliang Fan%
		\thanks{Chinese University of Hong Kong. Email: michaelqfan@gmail.com .}}
}

\date{\today}

\maketitle
} 



\vspace*{-0.2cm}
\begin{abstract}

\noindent 

In this review, we provide  practical guidance on some of  the main machine learning tools used in portfolio weight formation. This is not an exhaustive list, but a fraction of the ones used and have some statistical analysis behind it. All this research is essentially tied to precision matrix of excess asset returns. 
Our main point is that the techniques should be used in conjunction with outlined objective functions. In other words, there should be joint analysis of Machine Learning (ML) technique with the possible portfolio choice-objective functions in terms of test period Sharpe Ratio or returns. The ML method with the best objective function should provide the weight for portfolio formation. Empirically we analyze five time periods of interest, that are out-sample and show performance of some ML-Artificial Intelligence (AI) methods.  We see that nodewise regression with Global Minimum Variance portfolio based weights deliver  very good Sharpe Ratio and returns across five time periods in this century we analyze.  We cover three downturns, and 2 long term investment spans.


\noindent{\em Keywords:} covariance matrix estimation, factor models, shrinkage

\noindent {\em JEL Codes:} C40, C45, C58.
\end{abstract}

\newpage
\section{Introduction} \label{sec_intro}

In this article we investigate the practical aspects of ML-AI in portfolio advice given our experience. ML-AI is revolutionizing several aspects of life. One of these aspects is portfolio formation. Portfolio advice via ML-AI is preferred to give a fair share to retail investors who may be reluctant to use active investing. The belief is that ML-AI is objective and can protect against corrupt practices-bad and ill-intended advice in emerging markets. Another reason is that having research units may be extremely costly and ML-AI based methods can do do the same job with a fraction of money.

Having said all of that we will analyze a subset of ML-AI based methods that exists in econometrics literature. Our aim is to analyze the cases where the number of assets $p$ is larger than the   sample size $T$. A lot of times the number of assets may be more numerous than the sample size, either due to analyzing a large portfolio so that risk is minimized or availability of stable-stationary training data in limited time periods due to structural instability.

We start with the main component of portfolio investment advice in ML-AI methods we cover here. That will be a $p \times p$ dimensional precision matrix. Its estimation is key, since a simple inversion of sample covariance matrix fails due to singularity of sample covariance matrix when $p>T$. There is a great survey of several precision matrix estimation in \cite{crz2016}, and large discussion of several new methods in \cite{cd2025}. Among the methods we choose techniques that involve linear shrinkage estimator of covariance matrix of asset returns by \cite{lw2004}, and non-linear shrinkage estimator of covariance matrix of asset returns \cite{lw2017} with its single factor version. Then we analyze factor model based methods. These are the observed factor, thresholded covariance matrix of asset returns estimator of \cite{fan2011}, and its latent factor version in \cite{fan2013}. Both shrinkage and factor model based covariance matrix estimators are then inverted to get a precision matrix estimate. Next we analyze the nodewise regression based methods. These are well explained and developed by naive nodewise regression estimator found by \cite{mein2006}, and applied to portfolio variance estimate in \cite{caner2019}, and residual nodewise regression precision matrix estimate of  \cite{caner2022}. We do not cover recent deep learning based estimator of \cite{cd2025} since it can only estimate deep learn based estimator when $p<T$. We discuss assumptions and how it is done in a step by step algorithm in each method.

Next we introduce three objective functions-portfolio weight formation advice that is used in practice and literature. These are maximum Sharpe Ratio (MSR) portfolio that maximizes return/risk ratio for a portfolio, and Markowitz-mean-variance (MV) portfolio that maximizes return subject to a given variance, and the last one is Global Minimum Variance portfolio (GMV) that minimizes the variance of the portfolio only.

The main contribution and review of this article is joint analysis of precision matrix estimation techniques paired with portfolio weight formation of these three objective functions.  The portfolio advice should be the winner of each seven (number of precision matrix estimates) by three (portfolio weight objectives) matrix in terms of metric of choice. These metrics can be Sharpe Ratio, return of the portfolio, variance, and turnover.

In our empirical exercise we cover three stock downturn periods 2000-2003, 2008-2010, 2020-2023, and two long term periods 2000-2010, 2000-2019. The aim is to see how ML-AI perform in downturns of mixed nature, and also see some long term trends. We discover that naive-nodewise regression with GMV produces best Sharpe Ratio 4 out of 5 time periods and can beat $S\&P500$ too, unlike other ML-AI methods.

Section 2 covers precision matrix estimation. Section 3 covers shrinkage methods. Section 4 covers factor based methods. Section 5 analyzes nodewise regression methods. Section 6 is about portfolio formation. Section 7 obtains empirics, and Conclusion summarizes the findings.

\section{Precision Matrix}

In ML-AI based investing, portfolio weight formation, it is clear that we  use them with some financial guidance so that  better Sharpe Ratios, returns, small risk
can be obtained.  The key element to all these major decisions is the precision matrix of asset returns. To be specific, by asset returns we mean that the return of an asset over 3 month T-bill in USA example. So these are excess asset returns over a risk-free asset.  The current 3 month-T-bill rate as of July 11 2025  is 4.25\%. The return of Nvidia January 1 2025-July 11 2025 is 21.18\%, so excess asset return measured from January 1 2025-July 11 2025 for Nvidia is 16.93\%.

Why precision matrix is the key to al these issues? The main reason is that it forms the backbone of Sharpe Ratio, portfolio weight formulas, and the variance of the portfolio . What is the precision matrix on an intuitive level? Under normally distributed data, which we do not assume in theories, each cell in this matrix is the conditional correlation between different assets. For example, if we have a large portfolio including technology stocks including Nvidia and Palantir, the cell in the precision matrix that corresponds to Nvidia and Palantir tells us that correlation between two given all other stocks in our portfolio. If we believe for some reason there is no conditional correlation, we can set that number to zero. That implies that   there is no direct link between Nvidia and Palantir but still may be indirectly linked through other asset groups. On the other hand, if we put a nonzero conditional correlation between two stocks, it means there is a direct link between the two.

\subsection{High Dimensions}

It is possible that population covariance matrix is positive definite, hence invertible to get population precision matrix. However, when we try to get a sample estimate of the covariance matrix of asset returns it may not be invertible, when number of assets in our portfolio exceeds the time span of the analysis that we conduct. Let us denote the number of assets as $p$, and denote the time series as $T$. So the issue can stem when $p>T$. In this article, our interest centers on $p>T$ case, hence we will not consider techniques that are used but has $p \le T$. Another reason is that a lot of times in emerging markets and more developed markets due to uncertainty-structural changes, we estimate the models on small pockets of time, but have large number of assets to use.  Why large number of assets may be desirable? On reason is that variance of a large portfolio can be smaller than a small one, hence a risk averse investor, may work with that. 

We use the notation $Y: T \times p$ matrix of asset returns, where each row $y_t': 1 \times p$ shows the returns for all assets at time t.
Then a possible estimator for population covariance $\Sigma:= E ( y_t y_t'): p \times p$ is 
\[ S: = \frac{1}{T} Y'Y : p \times p.\]
We show that this is singular when $p>T$.  Note that sample covariance rank is $rank (Y'Y) = rank (Y)= T$, since $p>T$. The matrix is $p \times p$, hence it is less than full rank of $p$, hence singular and not invertible. Note that when $p\le T$, $S$ is invertible. So we cannot use $S^{-1}$ for precision matrix estimator.  We proceed then with 3 broad approaches to get precision matrix estimator in high dimensions. These are namely shrinkage based methods, factor model based ones, and nodewise regression based ones. These three broad categories are chosen, since they are used in not only in theory but practice as well with good empirical results. These 3 categories are in line with portfolio investing better, and can offer some statistical guarantees, hence they are not simple algorithms that can work in a given situation.

\section{Shrinkage Estimators}

In this area there are major theory contributions by \cite{lw2004,ledoit2008robust,ledoit2012,lw2017}.  Here we analyze linear-shrinkage, non-linear shrinkage and also talk about briefly about a variant of non-linear shrinkage estimator.  The idea of the shrinkage estimators are first we get a covariance matrix estimate and then invert it to get a precision matrix estimate. Hence the main estimator is a covariance matrix estimator and invertible, hence its inverse forms the precision matrix seamlessly.  Shrinkage covariance matrix estimators in $p>T$ cases have one major advantage over $S$, they are invertible. We start with linear shrinkage estimators.

\subsection{Linear Shrinkage}

Their main idea is to start thinking differently with a different sample covariance type of matrix. They assume that $y_1, \cdots, y_t, \cdots, y_T$ are independent across $t=1,\cdots, T$, and they have mean zero and covariance is $\Sigma$.
This is convex combination of Sample Covariance and Identity matrix. Why this formulation? This is needed since it is one way of getting a nonsingular covariance matrix estimator. Mainly they start with 
\[ \Sigma^* := \rho_1 I_p + \rho_2 S.\]

But $\rho_1, \rho_2$ parameters (ratios)  are arbitrary, so they choose these parameters that minimizes the expected quadratic loss $E \| \Sigma^* - \Sigma \|_{sF}^2$, where $\| .\|_{sF}$ is the scaled Frobenius norm of a given matrix, and $\| A \|_{sF} :=\sqrt{Tr (A A')/p}$. To get the optimal parameters in the new estimators, they realize that indeed, optimality depends on four parameters, since $\rho_1, \rho_2$ are ratios. Their Theorem 2.1 shows that 
\[ \rho_1= \frac{\beta^2}{\delta^2} \mu, \quad  \rho_2= \frac{\alpha^2}{\delta^2}.\]

These four scalars are defined as follows
\[ \mu:= Tr (\Sigma I_p)/p, \quad \alpha^2:= \| \Sigma - \mu I_p \|_{sF}^2, \quad \beta^2:= E \| S - \Sigma \|_{sF}^2, \quad  \delta^2:= E \| S - \mu I_p \|_{sF}^2.\]
They define $\mu I_p$ as the shrinkage target, and $\beta^2/\delta^2$ as the shrinkage intensity. Before estimators we setup the assumptions 1-3 in \cite{lw2004}. 
Note that we can have the following eigenvalue-eigenvector decomposition of $\Sigma:= \Gamma \Lambda \Gamma'$, in which $\Lambda: p \times p$ diagonal matrix that caries eigenvalues in the diagonal, $\Gamma$ is a rotation matrix that has the eigenvectors of $\Sigma$ as columns in $\Gamma$. 
 Define also the following transformed variable $Z:= \Gamma' Y$, which is described on p.375 of \cite{lw2004} as $p \times T$ matrix of iid observations on a system of $p$ uncorrelated random variables that spans the same space as the original system.  Let $(z_{11},\cdots, z_{p1})'$ as the first column of the matrix $Z$.

\begin{assum}\label{lwl1}
There exists a constant $K_1$ independent of $T$ such that $p/T \le K_1$.

\end{assum}

This assumption allows clearly $p>T$ since $K_1$ can be large. It does not allow exponentially growing $p$ in terms of $T$.

\begin{assum}\label{lwl2}
There exists a constant $K_2$ independent of $T$ such that  $\frac{1}{p} \sum_{i=1}^p E (z_{i1})^8 \le K_2$.

\end{assum}

This puts an eighth moment bound on asset returns, averaged over all assets.

\begin{assum}\label{lwl3}
\[ \lim_{ T \to \infty} \left(\frac{p^2}{ T^2} \right) \left(\frac{ \sum_{i,j,k,l \in Q} [ Cov (z_{i1}z_{j1}, z_{k1} z_{l1})]^2}{|Q|}\right) = 0,\]
where $Q$ denotes the set of all quadruples that are made of four distinct integers between 1 and $p$.

\end{assum}

$Cov(.,)$ shows the covariance between two random variables. Assumption states that products of uncorrelated random variables are uncorrelated in the limit. $|Q|$ is the cardinality of $Q$.

Define the following norm $\| A \|_T^2:= Tr (A A')/p$ with $A: p \times p $ matrix. Now we define the consistent estimators for  four parameters of the interest. 

\subsubsection{Algorithm to compute the estimator for precision matrix}

\begin{enumerate}

\item First 
\[ \hat{m}^2:= Tr (S I_p)/p,  \quad \hat{d}^2:=  
\| S - \hat{m} I_p \|_T^2.\]

\item Designate $y_t$ as the t th column of $Y: p \times T $, $t=1,\cdots, T$
\[ \bar{b}^2:= \frac{1}{T^2} \sum_{t=1}^T \| y_t y_t' - S \|_T^2, \hat{b}^2: = min (\bar{b}^2, \hat{d}^2),\]
and then set the remaining $\hat{a}^2 := \hat{d}^2 - \hat{b}^2$. Note that $\hat{m}^2, \hat{d}^2, \hat{b}^2, \hat{a}^2 $ are   consistent estimators in quadratic mean for $\mu^2, \delta^2, \beta^2, \alpha^2$ respectively.

\item Then the consistent estimator for $\Sigma^*$ in quadratic mean is 
\[ \hat{S}:= \frac{\hat{b}^2}{\hat{d}^2} \hat{m} I_p + \frac{\hat{a}^2}{\hat{d}^2} S.\]

\item Then the estimator for precision matrix is $\hat{S}^{-1}$.

\end{enumerate}

Theorem 3.2 of \cite{lw2004} shows under Assumptions \ref{lwl1}-\ref{lwl3}, $\hat{S}$ is a consistent estimator for $\Sigma^*$. But note also that $\Sigma^* \neq \Sigma$, hence this is not estimating the covariance matrix, but a linear combination of it. This is done since to allay invertibility issues, with $p>T$,  in sample plug-in version of $\Sigma$. Note that $\hat{S}$ is invertible and can consistently estimate $(\Sigma^*)^{-1}$. But note that precision matrix is $\Sigma^{-1}$. 

What are the advantages of linear shrinkage? First, it is model free and can be used in other fields of study. It is also a dense estimator, does not impose any zero links between  outcomes. 
One issue is that to solve the non-invertibility of the sample covariance, they propose a modified version of population covariance matrix and they consistently estimate that in $p>T$, not the actual covariance matrix.

\subsection{Non -Linear Shrinkage}

The paper of \cite{lw2017} starts with the main issue directly. The target is estimating the covariance matrix. Linear shrinkage above is a combination of sample covariance matrix and identity matrix, hence $O(1)$ parameters are needed to estimate instead of  $O(p^2)$. The non-linear shrinkage will estimate $O(p)$ parameters-degrees of freedom out of $O(p^2)$ data size if $p$ is close to T in terms of magnitude, with $T$ being the sample size and $p$ being the number of assets again.  The authors provide a rationale for estimating $O(p)$ degrees of freedom for $p^2$ parameters. They argue that estimating $O(1) $ parameters with data size $O(T^2)= O(p^2)$, when $p$ is close to $T$ in magnitude is too tight, and if the technique used $O(p^2)$ parameters, the technique will be loose.
The main technique here is pushing up the small eigenvalues up and larger ones down. The issue is how to choose the optimal shrinkage in a way that new covariance matrix estimate is more stable. The paper is mainly about choosing an estimator that minimizes the variance of a portfolio. The authors start with a loss function which is the  out-of-sample variance of a portfolio. The limit of this loss is provided and the limit depends on a shrinkage function and the authors find the  shrinkage function that optimizes the loss function and develop an estimate for the optimal shrinkage. We will try to illustrate their approach here.

The loss function depends on $\mu: p \times 1$ which is the target portfolio return and $\Sigma$ the population covariance matrix of returns. The optimal portfolio weight estimate is \begin{equation}
 \hat{w}:= \frac{\sqrt{\mu'\mu}}{\mu' \hat{\Sigma}^{-1} \mu} \hat{\Sigma}^{-1} \mu,\label{w1}
 \end{equation}
with $\hat{\Sigma}$ denoting a generic invertible estimator of the covariance matrix, and this is equation (3) of \cite{lw2017}. In Definition 1 of their paper, authors set the empirical loss function as the out-of-sample variance of the portfolio.
\begin{equation}
 L:= \hat{w}' \Sigma \hat{w}.\label{l1}
 \end{equation}

The next step will be the limit of the loss function and for that they will impose some assumptions.

\begin{assum}\label{lw-nl1}
Let $p/T \to c \in (0,1) \cup (1, + \infty)$, so $p$ cannot be close to T, but still it allows $p>T$.

\end{assum}

\begin{assum}\label{lw-nl2}

The population covariance matrix is nonrandom symmetric and positive definite matrix of dimension $p \times p$. The limiting spectral distribution exists and its well defined.

\end{assum}

\begin{assum}\label{lw-nl3}
Define $X$ as $T \times p$ matrix of iid random variables with mean zero, variance one, and finite 12th moment. $X$ is unobserved. Define $Y:= X \times \sqrt{\Sigma}$, which is observed. $\Sigma^{1/2}$ is symmetric positive definite by keeping the same eigenvectors as $\Sigma$ but recombining them with square roots of the population eigenvalues.

\end{assum}

Sample covariance is $S:= Y' Y/T$. It admits a spectral decomposition $S:= U \Lambda U'$, where $U: p \times p$ orthogonal matrix, that has eigenvectors columns of $U$. 

\begin{assum}\label{lw-nl4}

(i). The estimator for the covariance is defined as follows
\[ \hat{\Sigma}:= U \hat{\Delta} U',\]
by changing the diagonal of eigenvalues in the sample covariance matrix.

\[ \hat{\Delta}:= diag (\hat{\delta}(\lambda_1),\cdots \hat{\delta}(\lambda_p)),\]
and $\hat{\delta} (\lambda)$ is a function of eigenvalue of the sample covariance hence can depend on $S$. The eigenvalues are listed in increasing order, $\lambda_1$ is the minimum eigenvalue in the sample covariance matrix. 

(ii). Let $\hat{\delta}(x)\to \delta (x)$ almost surely for all x which is the support of F which is the limit of spectral distribution. $\delta(x)$ is continuously differentiable. Also this convergence is uniform over $\cup_{k=1}^{\kappa} [ a_k + \eta, b_k - \eta]$ for any small $\eta >0$.

(iii). Finally, for any small $\eta>0$, there exists a finite nonrandom constant $\hat{K}$  such that almost surely over the set $x \in \cup_{k=1}^{\kappa}
[ a_k - \eta, b_k + \eta]$ and $\hat{\delta}(x)$ is uniformly bounded by $\hat{K}$ from above and by $1/\hat{K}$ from below for all $T$ large enough. In the case $c>1$ there is the additional constraint $1/\hat{K} \le \hat{\delta} (0) \le \hat{K}$ for all $T$ large enough.

\end{assum}

Parts (ii)(iii) puts a strong structure on the shrinkage function, $\hat{\delta}(.)$  in high dimensional non-linear setting.

\begin{assum}\label{lw-nl5}
$\mu$  is distributed independently of $S$ and its distribution is rotation invariant.

\end{assum}

Under Assumptions \ref{lw-nl1}-\ref{lw-nl5}, Theorem 1 of \cite{lw2017} derives the following limit for their out-of-sample variance loss  at (\ref{l1}) 
\[ L \stackrel{a.s}{\to} g(x),\]
where 
\[ g(x):= \frac{\sum_{k=1}^{\kappa} \int_{a_k}^{b_k} \frac{d F (x)}{x | s(x)|^2 \delta (x)^2} + 1_{\{ c>1 \}} \frac{1}{c s(0) \delta(0)^2}}{\left[ \frac{d F(x)}{\delta (x)}
\right]^2},\] 
where for all $x \in (0, \infty)$ and also for $x=0$ in the case $c>1$, $s(x)$ is  defined as, with $s \in R \cup C^+$, with $C^+$ as the strict upper half of the complex plane,
\[ s(x):= - \left[ x - c \int \frac{\tau}{1+ \tau s} d H (\tau)
\right]^{-1},\]
where $H(.)$ represents the limit spectral distribution for the population covariance matrix and $F(.)$ is the limiting sample covariance spectral distribution. The main contribution here is the existence of a limit for the out-of-sample variance. But note that this is not the population variance in the limit for a portfolio. The limit is highly complicated, but crucially depends on the shrinkage function $\delta(x)$. The the issue becomes what is the optimal shrinkage function? The choice is the one that minimizes the limit above. It is given in their Theorem 2 by 
\[ d^*(x):= 1_{ \{ x >0 \}} \frac{1}{x |s (x)|^2} + 1_{ \{ c >1 \cap x=0 \}} \frac{1}{(c-1) s(0) },\]
with $x$ being in the support of $F$.   The issue then becomes how to choose a bona-fide estimator for $d^*(x)$. 
To choose an estimator the analyze the new modified sample covariance matrix estimator 
\[ \hat{S}:= U \hat{D} U',\]
where 
\[ \hat{D}:= diag (\hat{d} (\lambda_1), \cdots, \hat{d} (\lambda_p))',\]
where $\hat{d}(\lambda_i)$, $i=1,\cdots, p$ is a bona fide estimator for the optimal shrinkage function $d^*(x)$.

Via their Corollary 1 of \cite{lw2017} show that choice of $\hat{d} (\lambda_i)$, which will be defined below, when the $\hat{S}$ is used to form the sample loss in (\ref{w1}) meaning 
\[ \hat{L}:= \mu' \mu \frac{\mu' \hat{S}^{-1} \Sigma \hat{S}^{-1} \mu}{(\mu' \hat{S}^{-1} \mu)^2},\]
and then when we get the limit corresponding to that estimated loss that will correspond to the limit with the optimal shrinkage function. Note that the bona-fide shrinkage function estimator is 
\[ \hat{d} (\lambda_i):= 1_{ \{ \lambda_i >0 \} } \frac{1}{\lambda_i | \hat{s} (\lambda_i)|^2} + 1_{\{ p >T \cap \lambda_i =0\}} \frac{1}{(p/T -1) \hat{s}(0)},\]
with $\hat{s}(.)$ is the estimator for the Stieltjes transform $s(x)$ above.

\subsubsection{Algorithm to estimate the precision matrix}

\begin{enumerate}

\item Start with the sample covariance $S:= Y' Y/T$. It admits a spectral decomposition $S:= U \Lambda U'$, where $U: p \times p$ orthogonal matrix, that has eigenvectors columns of $U$. Obtain $U$.

\item  Estimate the Stieltjes transform for each non-negative eigenvalue, $ \hat{s}(\lambda_i)$, with $\lambda_i$ representing all positive eigenvalues in $\Lambda$ in step 1 above, and  for $p>T$ estimate also the Stieltjes transform for zero eigenvalues.

\item Form the optimal shrinkage function estimator
\[ \hat{d} (\lambda_i):= 1_{ \{ \lambda_i >0 \} } \frac{1}{\lambda_i | \hat{s} (\lambda_i)|^2} + 1_{\{ p >T \cap \lambda_i =0\}} \frac{1}{(p/T -1) \hat{s}(0)},\]
with $\hat{s}(.)$ is the estimator for the Stieltjes transform $s(x)$ above in Step 2.

\item  Form the diagonal matrix
 \[ \hat{D}:= diag (\hat{d} (\lambda_1), \cdots, \hat{d} (\lambda_p))'.\]

\item Form the optimal nonlinear shrinkage covariance estimator
\[ \hat{S}:= U \hat{D} U',\]

\item Invert $\hat{S}$ for the precision matrix estimator.

\end{enumerate}

In summary, the estimator is designed to minimize the limit of out-of-sample variance (with respect to a shrinkage function), but by using a precision matrix estimator based on a modified sample covariance matrix. The modification adjusts the eigenvalues in such a way that large ones are reduced and small ones are adjusted up. The bona-fide estimator for optimal shrinkage shows how it is done. So this is a specific precision matrix estimator tailored to minimizing the portfolio variance, converges to a limit but not to the limit with the variance with the population precision matrix. In other words, they  do not have  
\[ L \stackrel{p}{\to}  \frac{\mu'\mu}{\mu' \Sigma^{-1} \mu} \neq g(x).\]
Of course, the aim is to solve the sample empirical problem, hence the authors are not after a consistency result. Their method is model-free and dense hence has desirable features.

\subsection{Single Factor-Nonlinear Shrinkage}

This is the modified version  of non-linear shrinkage.  This is done so that we can better compare non-linear shrinkage with factor models. It is observed in empirical studies, this is one of the best methods to deliver small variance, good returns and Sharpe Ratio in U.S. market.  These are documented in \cite{lw2017} and \cite{caner2022}. They start with a single factor which is the  return on the equity weighted portfolio of stocks in the investment universe. Define the covariance matrix estimator $\hat{\Sigma}_f$ fitting this exact factor model. They post-multiply $Y$ by $\hat{\Sigma}_f^{-1/2}$. This operation removes the structure contained in the factor matrix. Then they apply non-linear shrinkage technique to $Y \hat{\Sigma}_f^{-1/2}$. Through Corollary 1 in non-linear shrinkage of \cite{lw2017}, they obtain the output as $\hat{\Sigma}_c$. The final covariance matrix estimator is obtained by transforming $\hat{\Sigma}_c$
\[ \hat{\Sigma}:= \hat{\Sigma}_f^{1/2} \hat{\Sigma}_c \hat{\Sigma}_f^{1/2}.\]
So in sum, this is exactly like non-linear shrinkage but they transform $Y$ by postmultiplying with $\hat{\Sigma}_f^{-1/2}$. In a sense they are scaling the outputs with factors. Then using the increasing the small eigenvalues and decrease the  large eigenvalues, and by Corollary 1 in \cite{lw2017} the end result is $\hat{\Sigma}_c$ not $\hat{S}$ as in the previous section. After that they modify this by factors again and obtain $\hat{\Sigma}$ to put back the factor structure.  So it is not model-free anymore, it is a hybrid between factor model and shrinkage. It is still a dense estimator, and there is no theory derived for this estimator in \cite{lw2017}.

\subsubsection{Algorithm for Single Factor Nonlinear Shrinkage}

\begin{enumerate}

\item Set up $\hat{\Sigma}_f$ through sample covariance. 

\item Then postmultiply $Y$ by $\hat{\Sigma}_f^{-1/2}$.

\item Form $\hat{\Sigma}_c$ by following steps 1-5 in the previous subsection for nonlinear shrinkage algorithm, and this matrix  should be the output of step 5.

\item  Transform covariance matrix to original by $\hat{\Sigma}= \hat{\Sigma}_f^{1/2} \hat{\Sigma}_c \hat{\Sigma}_f^{1/2}.$

\item The precision matrix estimate is $\hat{\Sigma}^{-1}$.

\end{enumerate}

Note that this is also dense, but possibly inconsistent estimator for precision matrix and its not estimating factor covariance matrix and feeding that into estimation of precision matrix. The estimation errors may carry over to the precision matrix.

\section{Factor Model Based Estimation}\label{factor}

There are two approaches here that we consider. One of them is with observed factors and the other is with unobserved factors.

\subsection{Observed Factors}

The paper of \cite{fan2011} starts with providing a high dimensional covariance matrix estimation. Then this is inverted to get a precision matrix estimate. Specifically they impose a factor model on understanding the asset returns. So this is not model free.  Let $p$ be the number of assets, and $T$ is the time period. 
\begin{equation}
y_{it} = b_i' f_t + u_{it},\label{f-1}
\end{equation}
where $i=1,\cdots, p, t=1, \cdots, T$, and $b_i: K \times 1$ factor loadings, $f_t: K \times 1$ are the observed factors. Then there is the imposition of 
$E (u_t | f_t) =0,$
with $u_t:=(u_{1t}, \cdots, u_{pt})'$ and $y_t:=(y_{1t},\cdots, y_{pt})': p \times 1$. Define the factor loadings matrix, $B:= ( b_1, \cdots, b_p)': p \times K$ matrix. Hence we can have (\ref{f-1}) in the vector form as 
\begin{equation}
y_t = B f_t + u_t.\label{f-2}
\end{equation}
Using (\ref{f-2}) the covariance matrix is well defined by the model
\begin{equation}
\Sigma = B \Sigma_f B' + \Sigma_u, \label{f-3}
\end{equation}
where $\Sigma$ is $p \times p$ matrix and $\Sigma_f: K \times K$ covariance matrix, and $\Sigma_u$ is the covariance matrix of errors. Note that aim of the article is to provide consistent estimation of $\Sigma$ and its inverse $\Sigma^{-1}$. The proposed estimator is 
\[ \hat{\Sigma} = \hat{B} \hat{\Sigma}_f \hat{B}' + \hat{\Sigma}_u^{Th},\]
and $\hat{B}: p \times K$ is the factor loading estimates by OLS and 
\[ \hat{B}= (\hat{b}_1, \cdots, \hat{b}_p)',\]
and 
\[ \hat{b}_i = argmin_{b_i} [\frac{1}{Tp} \sum_{t=1}^T \sum_{i=1}^p (y_{it} - b_i' f_t)^2].\]
The covariance matrix  estimator for factors is 
\[ \hat{\Sigma}_f= \frac{X X'}{T} - \frac{X 1_T 1_T' X'}{T^2},\]
where $X=(f_1, \cdots, f_T): K \times T$ matrix, and $1_T$ represents a vector of ones with $T$ dimension. $\hat{\Sigma}_u^{Th}$ is the thresholded covariance matrix of errors and will be explained below. The paper does not use covariance matrix estimator  for errors based on OLS. The reason is that since the article aims to provide 
consistent estimation in high dimensions with a dense OLS based covariance matrix of errors, that would not have been possible. Hence they need a sparse new covariance matrix estimator $\hat{\Sigma}_u^{Th}$. By thresholding the covariance matrix carefully, i.e. eliminating minor covariances according to a theoretical formula they achieve that. First, they  introduce $\hat{\Sigma}_u^{Th}$ and explain its properties. Denote the OLS residual 
\begin{equation}
\hat{u}_{it} = y_{it} - \hat{b}_i' f_t.\label{f-4}
\end{equation}  
Then $(i,j)$ th element of $\hat{\Sigma}_u^{Th}$ is denoted as $\hat{\sigma}_{ij}^{Th}$ 
\begin{equation}
\hat{\sigma}_{ij}^{Th}:= \hat{\sigma}_{ij} 1_{ \{ | \hat{\sigma}_{ij} | \ge \sqrt{\hat{\theta}_{ij}} \omega_T\}},\label{f-5}
\end{equation}
with $\hat{\sigma}_{ij}$ as the (i,j) th element of OLS based dense covariance matrix of errors which is 
\[ \hat{\Sigma}_u= \frac{1}{T} \sum_{t=1}^T \hat{u}_t \hat{u}_t', \quad p \times p,\]
with $\hat{u}_t= (\hat{u}_{1t}, \cdots, \hat{u}_{pt})'$ with $\hat{u}_{it}$ defined above in (\ref{f-4}). Then in thresholding 
\[ \hat{\theta}_{ij} = \frac{1}{T} \sum_{t=1}^T (\hat{u}_{it} \hat{u}_{jt} - \hat{\sigma}_{ij})^2,\]
and the rate is $\omega_T = O (\sqrt{log p/T})$. The specific $\hat{\theta}_{ij}$ and the rate $\omega_T$ will help in providing sparsity for the estimator and $\hat{\theta}_{ij}$ is the data dependent threshold level.  The convenience of factor model is, we can compute the precision matrix as long as covariance of factors is invertible, by Sherman-Morrison-Woodbury (SMW) formula
\[ \Sigma^{-1} = \Sigma_u^{-1} - \Sigma_u^{-1} B [ \Sigma_f^{-1} + B' \Sigma_u^{-1} B ]^{-1} B' \Sigma_u^{-1},\]
which gives rise to the following precision matrix estimator
\[ \hat{\Sigma}^{-1} = (\hat{\Sigma}_u^{Th})^{-1} -  (\hat{\Sigma}_u^{Th})^{-1} \hat{B} 
[ \hat{\Sigma}_f^{-1} + \hat{B}' (\hat{\Sigma}_u^{Th})^{-1} \hat{B} ]^{-1} \hat{B}' (\hat{\Sigma}_u^{Th})^{-1}.\]

We start with assumptions and one of the main results.

\begin{assum}\label{af-1}

(i). $u_t$ is stationary and ergodic such that $u_t$ has zero mean and covariance of $\Sigma_u$. $u_t$ is strong mixing with a coefficient $\alpha(t)$ with 
\[ \alpha (t) \le exp (-c t^{r_2}),\]
with $r_2>0$, $c>0$ positive constants for all $t$ which is a positive integer. 

(ii). There exists constants $c_1, c_2$ all positive constants with 
\[ c_1 < Eigmin(\Sigma_u) \le Eigmax (\Sigma_u) < c_2,\]
with 
\[ c_1 < var (u_{it} u_{jt}) < c_2,\]
for the variance var(.) for all $1 \le i \le p, 1\le j \le p$.

(iii). There exists $r_1>0, b_1>0$ such that for any $s>0$ and $1\le i \le p, 1\le t \le T$
\[ P (| u_{it} | >s) \le exp (-s/b_1)^{r_1}.\]

\end{assum}

This assumption allows time series correlation but in (ii) assumption is restricting the maximum eigenvalue to be constant in a large $p \times p$ matrix, a bit restrictive for noisy financial markets.

\begin{assum}\label{af-2}
There exist positive sequences $k_1= o(1)$, and $k_2=o(1)$ and a constant $M>0$ and a positive sequence $a_T=o(1)$ with a constant $C >M$ 

(i). \[ P [ \max_{1 \le i \le p} \frac{1}{T} \sum_{t=1}^T | u_{it} - \hat{u}_{it} |^2 > C a_T^2] \le O (k_1),\]

(ii). \[ P [ \max_{1 \le i \le p} \max_{1 \le t \le T } | u_{it} - \hat{u}_{it} | > C ] \le O (k_1).\]

\end{assum}

This Assumption \ref{af-2} is a high level assumption and it is not clear it will be holding all cases. 

\begin{assum}\label{af-3}

(i). $f_t$ is stationary and ergodic and strong mixing with the same strong mixing coefficient as $u_t$.

(ii). $u_t$ and $f_t$ are independent.

\end{assum}

Assumption \ref{af-3} allows $f_t$ to have a time series correlation, but there should be independence from the errors. 

\begin{assum}\label{af-4}

(i). There exists a constant $M >0$ such that for all $i,j,t$ $E y_{it}^2 < M$, $E f_{it}^2 < M$ and $| b_{ij} | < M$.

(ii). There exists  a constant $r_3>0$ with  $\frac{1}{r_3}+ \frac{1}{r_2} >1$ and $b_2 > 0$ such that for any $s>0$ and $i \le K $
\[ P ( | f_{it} | > s) \le exp (-(s/b_2)^{r_3})\]

\end{assum}

\begin{assum}\label{af-5}
There exist a constant $C>0$ such that $Eigmin (\Sigma_f ) > C >0$.

The last two assumptions provide a basis for Bernstein type inequality.

\end{assum}

Let $\gamma_2^{-1} = 1.5 r_1^{-1} + 1.5 r_3^{-1} + r_2^{-1}$

\begin{assum}\label{af-6}

Suppose $K =o(p)$  and $K^4 (log p)^2 = o(T)$, and $(log p)^{2/\gamma_2}-1 = o(T)$.

\end{assum}

The last assumption shows that $K$ can grow but should be slower than $p, T$ and also we can have $p>T$. The following result is a combination Theorem 3.1(ii), and Theorem 3.2 (ii) in \cite{fan2011}. Let $\| .\|$ be an $l_2$ operator norm for matrices, and $m_T$ be the maximum number of nonzero cells across rows of the covariance matrix of errors $\Sigma_u$.

Under Assumptions \ref{af-1}-\ref{af-6} with $m_T K \sqrt{log p/T} \to 0$ and $log T/p \to 0$

 \[ \| (\hat{\Sigma}_u^{Th})^{-1} - \Sigma_u^{-1} \| = O_p (m_T K \sqrt{log p/T} ) = o_p (1).\]

\[ \| \hat{\Sigma}^{-1} - \Sigma^{-1} \| = O_p ( m_T K \sqrt{log p/T}) = o_p (1).\]

First we see that precision matrix estimator for errors is consistently estimable even when $p>T $ in (i), and then we see that also precision matrix of outcomes is consistently estimable with $p>T$. These two results are remarkable since it shows high dimensional consistent estimation is possible with factor model imposed on the asset returns. Hence it is possible to merge financial literature-practice-statistics.

\subsubsection{ Algorithm to run Factor Model based Precision Matrix estimate}

\begin{enumerate}

\item Run OLS regression $y_{i}=(y_{i1}, \cdots, y_{iT}): T \times 1$ on $f_t:K \times 1$ to get $\hat{b}_i: K \times 1: (XX')^{-1} X y_i$

\item Obtain residuals $\hat{u}_{it} = y_{it} - \hat{b}_i' f_t$,  and obtain $\hat{\sigma}_{ij}$ as the (i,j) th element of the estimated sample covariance matrix of errors.
\[ \hat{\Sigma}_u= \frac{1}{T} \sum_{t=1}^T \hat{u}_t \hat{u}_t', \quad p \times p.\]

\item Then for the thresholded covariance matrix estimate of errors, setup \[ \hat{\theta}_{ij} = \frac{1}{T} \sum_{t=1}^T (\hat{u}_{it} \hat{u}_{jt} - \hat{\sigma}_{ij})^2\]

\item Form (\ref{f-5}) for each $i,j$ element of the thresholded covariance matrix estimate of errors, can use $\omega_T = 0.1 K \sqrt{log p/T}$, p.3335 of \cite{fan2011} which gives good results. Obtain $\hat{\Sigma}_u^{Th}$.

\item Obtain the covariance matrix of factors estimator 
\[ \hat{\Sigma}_f= \frac{X X'}{T} - \frac{X 1_T 1_T' X'}{T^2}.\]

\item Obtain the precision matrix estimator for returns by SMW formula.
\[ \hat{\Sigma}^{-1} = (\hat{\Sigma}_u^{Th})^{-1} -  (\hat{\Sigma}_u^{Th})^{-1} \hat{B} 
[ \hat{\Sigma}_f^{-1} + \hat{B}' (\hat{\Sigma}_u^{Th})^{-1} \hat{B} ]^{-1} \hat{B}' (\hat{\Sigma}_u^{Th})^{-1}.\]

\end{enumerate}

In sum, \cite{fan2011} is estimating the precision matrix of asset returns with an imposed factor model estimation technique, and since the covariance matrix of errors 
are the main roadblock to this consistent estimation, they impose sparsity on that. That is a minimal assumption, since there is no imposition of sparsity on the asset return covariance. The main issue with this method it is not model estimation free. For the precision matrix of returns instead of using the actual data and estimate that, they use a predicted version through observed factors.
If the factor model fit is incorrect all the precision matrix estimation may be incorrect.

\subsection{Principal Orthogonal Complement Thresholding Estimator}\label{poet}

\cite{fan2013} introduces a hidden factor model version of the observed factor model of \cite{fan2011}.
The paper introduces POET (Principal Orthogonal complEment Thresholding) estimator and this is one of the most widely used precision matrix estimators. It is widely used in financial applications as a covariance matrix estimator of asset returns. So it is  specific to finance as in the previous section. This is proposed by \cite{fan2013}. The structure is very similar to the previous section but the main difference is that factors are latent. This may  be due to behavioral factors-sentiments not observed and still affecting the asset returns. Using the same notation as in the previous subsection

\begin{equation}
y_t = B f_t + u_t,\label{poet-1}
\end{equation}
where $f_t$ are latent factors now. The covariance matrix of asset returns is 
\[ \Sigma= B \Sigma_f B' + \Sigma_u,\]
and precision matrix is $\Sigma^{-1}$. One other difference is that $\Sigma_u$ is approximately sparse, not exactly sparse as in the previous section. The sparsity index is defined as follows
\[ m_p:= \max_{1 \le i \le p} \sum_{1 \le j \le p}  | \sigma_{u,ij} |^{q},\]
with $q \in [ 0,1)$ and $ \sigma_{u,ij}$ is the $(i,j)$ th element of matrix $\Sigma_u$. In this sense when $q=0$ we have the exact sparsity as in the previous section, with $q \to 1$ we have approximate sparsity. $m_p  \to \infty$ when $p \to \infty$. They cast the estimation  as constrained least squares in Section 2.3 of \cite{fan2013} and provide the following estimator for the covariance estimator for the asset returns

\[ \hat{\Sigma} = \hat{B} \hat{B}' + \hat{\Sigma}_u^{T},\]
with $\hat{B}= T^{-1} Y \hat{F}$ which is the $p \times K$ matrix of factor loadings, and $K$ are the number of hidden factors, with $Y: p \times T$ matrix of asset returns, and each row of $Y$ represent a specific asset return across different periods of time, and each column represents  at specific time $t$, all asset returns. Note that estimate of covariance matrix of factors are normalized to identity matrix with dimension K.
We designate $\hat{F}: T \times K $ estimated factor matrix, and 
\[ \hat{F} = argmax_F tr [ (I_T - T^{-1} F F') Y'Y]
,\]
and the solution is that $\hat{F}/T^{1/2}$ are the matrix that is formed by the eigenvectors correspond to the largest eigenvalues of $Y'Y$. Then the thresholded covariance matrix of errors, $\hat{\Sigma}_u^T$ is considered. First, set the residual
\[ \hat{u}_{it}= y_{it} - \hat{b}_i' \hat{f}_t,\]
with $\hat{b}_i$ are the columns of $\hat{B}$ and $\hat{f}_t: K \times 1$ and they are the columns of $\hat{F}$. Then 
\[ \hat{\sigma}_{ij} = \frac{1}{T} \sum_{t=1}^T \hat{u}_{it} \hat{u}_{jt}.\]  
Then define the thresholding variable
\[ \hat{\theta}_{ij}:= \frac{1}{T} \sum_{t=1}^T (\hat{u}_{it} \hat{u}_{jt} - \hat{\sigma}_{ij})^2,\]
and define the rate 
\[ \omega_T:= \frac{1}{\sqrt{p}} + \sqrt{\frac{log p}{T}},\]
and see that $\omega_T$ has extra $1/\sqrt{p}$ compared to observed factor setup. The adaptive threshold part is
\[ \tau_{ij} = C \sqrt{\hat{\theta}_{ij}} \omega_T,\]
where $C>0$ is a positive constant. The main diagonal element of $\hat{\Sigma}_u^T$ is just $\hat{\sigma}_{ii}$, which is the sample average of squared residuals, and the off-diagonal elements are 
\[ \hat{\sigma}_{ij}^T:= \hat{\sigma}_{ij} 1_{ \{ | \hat{\sigma}_{ij} | \ge \tau_{ij}
\} },\]
which shows that covariances are cut to zero above a threshold level $\tau_{ij}$. Threshold level is not arbitrary, and designed by econometric theory.
By using SMR formula in the last section they provide the following estimated precision matrix
\[ \hat{\Sigma}^{-1} = [\hat{\Sigma}_u^T]^{-1} - [\hat{\Sigma}_u^T]^{-1} \hat{B} 
( I_K + \hat{B}' [\hat{\Sigma}_u^T]^{-1} \hat{B})^{-1} \hat{B}' [\hat{\Sigma}_u^T]^{-1}.\]
There in the above formula we use a fixed K. However, in practice we need to estimate unknown number of hidden factors. The authors use Bai and Ng (2002) 
formula
\[ \hat{K}:= argmin_{1 \le K_1 \le M} log \{ \frac{1}{pT} \| Y - T^{-1} \hat{F}_{K_1} \hat{F}_{K_1} \|_F^2 + K_1 g(p,T),
\} ,\]
with 
\[ g(p,T):= \frac{p+T}{pT} log min (p,T),\]
where $\hat{F}_{K_1}: T \times K_1$ matrix whose columns are $\sqrt{T}$ times the eigenvectors correspond to the largest $K_1$ eigenvalues of $T \times T$ matrix $Y'Y$, and $\|.\|_F$ is the Frobenius norm of the matrix. $M$ is an prespecified upper bound on number of hidden factors.

Now we provide assumptions.

\begin{assum}\label{poet1}
All the eigenvalues of $K \times K$ matrix $B'B/p$ are bounded away from zero and infinity as $p \to \infty$.

\end{assum}

This assumption imposes that the first $K$ eigenvalues of $\Sigma_y$ grow at rate $O(p)$. This means that all factors are pervasive hence strong.

\begin{assum}\label{poet2}
(i). $\{ u_t, f_t\}$ is strictly stationary, and $E u_{it}=0, E u_{it} f_{jt}=0$ for all $ 1 \le i \le p$, $1 \le j \le K$ and $1 \le t \le T$.

(ii). There exist constants $c_1, c_2>0$ such that $Eigmin (\Sigma_u) \ge c_1$, $\| \Sigma_u \|_1 < c_2$ and $\min_{1 \le i \le p} \min_{1 \le j \le p}
var (u_{it}, u_{jt}) > c_1$.

(iii). There exist $r_1, r_2 >0$ and $b_1 > 0, b_2 >0$ such that for any $s>0$, $i\le p$, $j \le K$
\[ P [ | u_{it} | >s] \le exp (-(s/b_1)^{r_1}),\]
\[ P | f_{jt} | >s] \le exp (-(s/b_2)^{r_2}).\]

\end{assum}
Note that (i) is pretty standard, but in (ii) $l_1$ matrix norm being bounded for covariance matrix of errors is strong assumption, (iii) is also standard and gives rise to exponentially declining tails and is a sufficient condition for maximal inequality in time series.

\begin{assum}\label{poet3}
$\{ f_t, u_t\}$ are strong mixing with the same mixing coefficient $\alpha (T) \le exp (- C T^{r_3})$ with $r_3>0$ and $3 r_1^{-1} + 1.5 r_2^{-1} + r_3^{-1} >1$.
\end{assum}

This assumption is needed for maximal inequality as well.

\begin{assum}\label{poet4}
There exist $M>0$ such that for all $1 \le i \le p$, $1 \le t \le T$, $1 \le s \le T$

(i). $\| b_i \|_{\infty} < M$.

(ii). $ E [ p^{-1/2} (u_s' u_t - E u_s' u_t)]^4 < M$.

(iii). $ E \| p^{-1/2} \sum_{i=1}^p b_i u_{it} \|^4 < M$.

\end{assum}

These are moment bound assumptions. 

\begin{assum}\label{poet5}
(i). Suppose $log p = o(T^{1/16}), T = o(p^2)$. 

(ii). $m_p \omega_T^{1-q} \to 0$. 

\end{assum}

The first (i) allows  exponential growth in T for $p$, and also puts an upper bound on growth of T, (ii) is a sparsity condition showing with exact sparsity $q=0$, the convergence is faster. The following is Theorem 3.2 of \cite{fan2013}, and uses the spectral norm for the matrices.

Under Assumptions \ref{poet1}-\ref{poet5}, POET estimator satisfies

\[ \| \hat{\Sigma}^{-1} - \Sigma^{-1} \|  = O_p (m_p \omega_T^{1-q}) = o_p(1).\]

Remarks. 1. $\hat{\Sigma}^{-1}$ uses data dependent $\hat{K}$ in theorem, and it is described before the theorem.

2. The rate of convergence $m_p \omega_T^{1- q}$ depends on sparsity $m_p$, and the rate of convergence for sup norm for $\hat{\Sigma}_y$ which is given in \cite{fan2013} as
\[ \| \hat{\Sigma} - \Sigma\|_{\infty} = O_p (\omega_T).\]

3. $K$ is fixed and $\omega_T$ has an additional rate here $1/\sqrt{p}$ compared to known factor case of \cite{fan2011}. This also shows that with fixed $p$ there is no consistency.

4. The paper allows $p>T$ consistency when both terms $p, T$ grow. Although consistency is more difficult to achieve with a less sparse setup when $q \to 1$
due to $m_p \omega_T^{1-q} \to 0$ condition.

\subsubsection{ Algorithm to run POET Precision Matrix estimate}

\begin{enumerate}

\item Select the unknown number of factors
\[ \hat{K}:= argmin_{1 \le K_1 \le M} log \{ \frac{1}{pT} \| Y - T^{-1} \hat{F}_{K_1} \hat{F}_{K_1} \|_F^2 + K_1 g(p,T),
\} ,\]
with 
\[ g(p,T):= \frac{p+T}{pT} log min (p,T),\]
where $\hat{F}_{K_1}: T \times K_1$ matrix whose columns are $\sqrt{T}$ times the eigenvectors correspond to the largest $K_1$ eigenvalues of $T \times T$ matrix $Y'Y$, and $\|.\|_F$ is the Frobenius norm of the matrix. $M$ is an prespecified upper bound on number of hidden factors.

\item Set $\hat{K}= K^*$.
Obtain the scaled estimated factor loadings corresponding to $K^*$ no of factors: $\hat{F}/T^{1/2}$ which is the matrix that is formed by the eigenvectors correspond to the largest eigenvalues of $Y'Y$.

\item Each row of $\hat{F}$ forms $\hat{f}_t$, $t=1,\cdots, T$.

\item Form the factor loading matrix estimate$\hat{B}= T^{-1} Y \hat{F}$ which is the $p \times K^*$ matrix of factor loadings.

\item Residuals are $\hat{u}_{it} = y_{it} - \hat{b}_i' f_t$, where $\hat{b}_i'$ are the rows of factor loading matrix $\hat{B}$.

\item Obtain $\hat{\sigma}_{ij}$ as the (i,j) th element of the estimated sample covariance matrix of errors.
\[ \hat{\sigma}_{ij}= \frac{1}{T} \sum_{t=1}^T \hat{u}_{it} \hat{u}_{jt}, \quad p \times p.\]

\item Then for the thresholded covariance matrix estimate of errors, setup \[ \hat{\theta}_{ij} = \frac{1}{T} \sum_{t=1}^T (\hat{u}_{it} \hat{u}_{jt} - \hat{\sigma}_{ij})^2\]

\item Set up the rate that will be used in thresholding:
\[ \omega_T:= \frac{1}{\sqrt{p}} + \sqrt{\frac{log p}{T}}\]

\item Form the thresholding variable
\[ \tau_{ij} = 0.5 \sqrt{\hat{\theta}_{ij}} \omega_T.\]

\item Form the estimate, thresholded covariance matrix of errors.
The main diagonal element of $\hat{\Sigma}_u^T$ is just $\hat{\sigma}_{ii}$, which is the sample average of squared residuals, and the off-diagonal elements are 
\[ \hat{\sigma}_{ij}^T:= \hat{\sigma}_{ij} 1_{ \{ | \hat{\sigma}_{ij} | \ge \tau_{ij}
\} },\]
which shows that covariances are cut to zero above a threshold level $\tau_{ij}$.

\item Obtain the precision matrix estimate
\[ \hat{\Sigma}^{-1} = [\hat{\Sigma}_u^T]^{-1} - [\hat{\Sigma}_u^T]^{-1} \hat{B} 
( I_K + \hat{B}' [\hat{\Sigma}_u^T]^{-1} \hat{B})^{-1} \hat{B}' [\hat{\Sigma}_u^T]^{-1}.\]

\end{enumerate}

In sum, the POET estimator is one of the most widely used estimators in academia and practice. It is consistent even with a less sparse (semi dense) setup for $\Sigma_u$. The estimator is not model-free and not consistent when $p$ is small. For small portfolio estimation problems, that may not be ideal choice.

\section{Nodewise Regression}

In this section we show nodewise regression based approaches. The first one is naive-nodewise regression, and the second one is a residual based nodewise regression.

\subsection{Naive Nodewise Regression}

This is mainly a method developed by \cite{mein2006}, and applied to financial econometrics-variance-risk estimation in large portfolios by \cite{caner2019}. One of the  main ideas of nodewise regression is to use a matrix algebraic exact closed form solution for each row of the precision matrix, and then stack all rows to get a precision matrix formula. In each row, a specific main diagonal term formula and off-diagonal term vector is derived. Then lasso estimation is used to estimate this population based formula, since it is a high dimensional problem. Next we show how these are done here. Define $y_t:  p \times 1$ vector of random variables, and for example it  can be vector of excess asset returns (returns which are above a risk free 3 month T-bill rate). Set $E y_t = \mu$, for $t=1,\cdots, n$, and 
\[ \Sigma:= E (y_t - \mu ) (y_t - \mu)',\]
which is a $p \times p$ matrix. The precision matrix is defined as $\Theta:= \Sigma^{-1}$. $\Sigma$ is of full rank, so it is invertible. Define 
\[ \bar{y}:= \frac{1}{n} \sum_{t=1}^n y_t.\]
So $t=1,\cdots,n$ represents time span, and $j=1, \cdots, p$ is the number of outcome variables (i.e. it can be number of assets in our portfolio). Let $y_{t,j}$ as the $j$ th outcome and time $t$.
Now we start introducing the formulae for the row of the precision matrix. Each row consists of the main diagonal term (i.e. in $j$ th row, $j$ th element is the main diagonal term) and the off-diagonal terms. Next two equations are also used in from \cite{caner2019}, \cite{canerkock2018}. The main diagonal term is
\begin{equation}
 \Theta_{j,j}:= [ \Sigma_{j,j} - \Sigma_{j,-j} \Sigma_{-j,-j}^{-1} \Sigma_{-j,j}]^{-1},\label{nw1}
 \end{equation}
where $\Sigma_{-j,-j}$ is the $p-1 \times p-1$ sub matrix of $\Sigma$ where $j$ th row and column are deleted. We define $\Sigma_{j,-j}$ as the $ j$ th row  of $\Sigma$ with $j$ th element is deleted. In the same way, $\Sigma_{-j,j}$ is the $j$ th column of $\Sigma$ with $j$ th element deleted. The off-diagonal terms in the $j$ 
\begin{equation}
 \Theta_{j,-j}:= - \Theta_{j,j} \Sigma_{j,-j} \Sigma_{-j,-j}^{-1}.\label{nw2}
 \end{equation}
Then we will relate the last two equations to linear regression. Define 
\[ y_{t,j}^*= y_{t,j} - \bar{y}.\]
Denote the vector $y_{t,-j}^*$ ($p-1 \times 1$) as the vector of all demeaned outcomes except for the $j$ th one. Then \cite{chang2019}, \cite{caner2019}  show with time series data (to be specified in assumptions below) the linear regression model is, for $j=1,\cdots,p$ 
\[ y_{t,j}^* = y_{t-j}^{*'} \gamma_j + \eta_{t,j},\]
which $\eta_{t,j}$ is the error term and  we relate the coefficient vector to precision matrix formation in the following way
\begin{equation}
\gamma_j = \Sigma_{-j, -j}^{-1} \Sigma_{-j,j}.\label{nw3}
\end{equation}
Using (\ref{nw3}) we can write (\ref{nw2}) as
\begin{equation}
\Theta_{j,-j} = - \Theta_{j,j} \gamma_j'.\label{nw4}
\end{equation}
 So if we figure out $\Theta_{j,j}$ in terms of $\gamma_j$ in (\ref{nw4}) above we will relate the matrix inversion to linear regression. By recasting (\ref{nw1}) using (\ref{nw3})
 \begin{equation}
 \Theta_{j,j}= ( \Sigma_{j,j} - \Sigma_{j,-j} \gamma_j)^{-1}.\label{nw5}
 \end{equation}
 We now give an example by analyzing the formula for the first row of the precision matrix, the other rows are treated in the same way. At $j=1$ using (\ref{nw4})(\ref{nw5})
 \[ \Theta_1 = \left( \frac{1}{\Sigma_{11} - \Sigma_{1,-1} \gamma_1}, \frac{-\gamma_1'}{\Sigma_{11} - \Sigma_{1,-1} \gamma_1}
 \right)'.\]
 To make things easier  in estimation subsequently define  the scalar for each $j=1,\cdots,p$ which is the reciprocal of the main diagonal terms in the precision matrix
 \[ \tau_j^2:= \frac{1}{\Theta_{j,j}} = \frac{1}{ ( \Sigma_{j,j} - \Sigma_{j,-j} \gamma_j)^{-1}}.\]
 
  Repeating this process for all $j=2,\cdots,p$ provides the precision matrix formula related to regression coefficient vector $\gamma_j$, resulting in 
 \[ \Theta = T^{-2} C,\]
 where $T^{-2}:= diag ( \tau_1^{-2}, \cdots, \tau_p^{-2})$ which is diagonal matrix with $p \times p$ dimension. $C$ is defined as
 \[ C:= \left[ \begin{array}{cccc}
 1 & -\gamma_{1,2} & \cdots & -\gamma_{1,p} \\
 -\gamma_{2,1} & 1 & \cdots & -\gamma_{2,p} \\
 \vdots & \vdots & \vdots & \vdots \\
 - \gamma_{p,1} & - \gamma_{p,2} & \cdots & 1
 \end{array}
  \right].\]
 
\subsubsection{Estimation}

Next we show how to estimate the precision matrix. Estimation of $\gamma_j$ will be done with lasso so that high dimensional consistency is achievable. In a sense, this is by imposing exact sparsity on the precision matrix and then proceeding with estimation. First, we set 
\[ S_j:=\{ k: \gamma_{j,k} \neq 0 \},\]
and define $s_j= |S_j|$ be the cardinality of the set $S_j$. Note that $S_j$ is the index of nonzero cells in $\gamma_j$ vector and $s_j$ is the number of nonzero cells in $\gamma_j$. Define 
\[ \bar{s}:= \max_{1 \le j \le p} s_j ,\]
as the maximum number of nonzero cells across the rows of the precision matrix. 
Note that sparsity of the factor models in \cite{fan2011}, \cite{fan2013} is different and it is on the covariance matrix of errors. The advantage of the precision matrix sparsity assumption is that covariance matrix can be dense but precision matrix can be sparse, such as Toeplitz matrices, and examples can be seen in \cite{caner2022}. Why this is important? Under the assumption of normality, the cells in the precision matrix are the conditional correlation between different asset returns. This is a key statement, since given all other asset returns, we form a relation between two specific asset returns in each cell. This is unlike unconditional covariance which does not carry information about other assets in the stock market. 

Since sparsity is imposed, a sparse estimation technique like lasso is employed for $\gamma_j$ estimation. Specifically for $ j=1,\cdots,p$
\begin{equation}
 \hat{\gamma}_j:= argmin_{\gamma_j \in R^{p-1}} 
[ \| y_j^* - Y_{-j}^* \gamma_j \|_n^2 + 2 \lambda_j \| \gamma_j \|_1],
\label{nw6}
\end{equation}
where $y_j^*: n \times 1$ vector of time series variables for the outcome variable $j$, and $Y_{-j}^*: n \times p-1$ matrix of all outcome variables except the $j$ th one. $\lambda_j$ is the tuning parameter and $\lambda_j>0$. The idea is to estimate $C$ and $T^{-2}$ matrices by using $\hat{\gamma}_j$ in (\ref{nw6}). In that sense we define
\begin{eqnarray}
\hat{C}:= \left[ \begin{array}{cccc}
1 & -\hat{\gamma}_{1,2} & \cdots & -\hat{\gamma}_{1,p} \\
-\hat{\gamma}_{2,1} & 1 & \cdots & -\hat{\gamma}_{2,p}\\
\vdots & \vdots & \vdots & \vdots \\
-\hat{\gamma}_{p,1} & -\hat{\gamma}_{p,2} & \cdots & 1 
\end{array}
\right],\label{nw7}
\end{eqnarray}
and 
\[ \hat{T}^{-2}:=  diag (\hat{\tau}_1^{-2},\cdots, \hat{\tau}_p^{-2}),\]
which is a $p \times p$ diagonal matrix with 
\begin{equation}
\hat{\tau}_j^2:= \frac{ \| y_j^* - Y_{-j}^* \hat{\gamma}_j \|_2^2}{n} + \lambda_j \| \hat{\gamma}_j \|_1.\label{nw8}
\end{equation}
Then define the estimator of inverse
is 
\[ \hat{\Theta}:= \hat{T}^{-2} \hat{C}.\]
Note $\hat{\Theta}$ is not symmetric and we can make that symmetric by $\hat{\Theta}_{sym}:= \frac{\hat{\Theta}+ \hat{\Theta}'}{2}$.  For theory $\hat{\Theta}$ will be consistent, and extending to $\hat{\Theta}_{sym}$ is simple, and not pursued. The tuning parameter is chosen by minimizing Generalized Information Criterion (GIC) as proposed in \cite{ft2013} with $\hat{\sigma}_{\lambda_j}^2:= \| y_j^* - Y_{-j}^* \hat{\gamma}_j \|_2^2/n$ and $|\hat{S}_{\lambda_j}|$ is the cardinality estimated by using a specific $\lambda_j$. GIC is 
\[ GIC (\lambda_j) := log (\hat{\sigma}_{\lambda_j}^2) + | \hat{S}_{\lambda_j}| \frac{log p}{n} log (log (n)).\]

We provide the assumptions that will be crucial in obtaining consistency of precision matrix estimate. These are standard and used in \cite{caner2019}, \cite{chang2019}. They show time series nature of the problem, and a sparsity condition. The key is Assumption \ref{nw4} which is the rate of convergence of precision matrix. So it allows for $p>n$ consistency, which is a key result in \cite{caner2019}.

\begin{assum}\label{nw1}

The $n \times p$ matrix of excess asset returns $y^*$ has strictly stationary $\beta$ mixing rows with $\beta$ mixing coefficients satisfying $\beta_k \le exp (- K_1 k^{\xi_1})$ for any positive constants $K_1, k, \xi_1$ that are independent of $n,p$.

\end{assum}

\begin{assum}\label{nw2}

The smallest eigenvalue of $\Sigma$ is $\Lambda_{min}$ is strictly positive and uniformly bounded away from zero. The maximum eigenvalue of $\Sigma$ is uniformly bounded away from infinity.

\end{assum}

\begin{assum}\label{nw3}
There exists constants $K_2>0, K_3>1$, $0 < \Xi_2  \le 2, 0 < \Xi_3 \le 2$ that are independent of $p,n$ such that 
\[ \max_{1 \le j \le p} E [exp (K_2 | y_{t,j}|^{\Xi_2})] \le K_3,\]
\[ \max_{1 \le j \le p} E [exp (K_3 | \eta_{t,j}|^{\Xi_3})] \le K_3,\]

\end{assum}

\begin{assum}\label{nw4}
The following condition holds 
\[ \bar{s} \sqrt{logp/n} \to 0.\]

\end{assum}

\subsubsection{Algorithm for Naive Nodewise Regression}

\begin{enumerate} 

\item Select a grid for $\lambda_j$ for each $j$. We can have $\Lambda_j :=\{ \lambda_{j1}, \lambda_{j2},\cdots, \lambda_{jq}\}$ with $q$ as the number of possible $\lambda_j$ choices.
 
\item Estimate $\hat{\gamma}_j$ for a given $\lambda_j$ in the grid in step 1 by (\ref{nw6}). Pair each $\lambda_j$ choice with corresponding $\hat{\gamma}_j$.

\item  Repeat this for all $\gamma_j \in \Lambda_j$ to get the optimal $\lambda_j^*$ that minimizes GIC above.

\item Then for that $\lambda_j^*$ find the corresponding $\hat{\gamma}_j$ in Step 2 and denote it as $\hat{\gamma}_j^*$.

\item Form $\hat{C}$ and $\hat{T}^{-2}$ given $\hat{\gamma}_j^*$ using (\ref{nw7})(\ref{nw8}) and obtain $\hat{\Theta}:= \hat{T}^{-2} \hat{C}$.

\end{enumerate}

This approach is called nodewise regression and tied to neighborhood selection in graph models as shown in Section 2.3 of \cite{caner2019} or \cite{mein2006}.
Next we show how nodewise regression can be employed in large portfolios. We start with our assumptions. These are standard in the literature.

The article by \cite{caner2019} under Assumptions \ref{nw1}-\ref{nw4} uses nodewise regression  result $\hat{\Theta}$ consistently estimates $\hat{\Theta}$ in high dimensions in time series data even when $p>n$ by \cite{chang2019}, and both growing.  \cite{caner2019} uses this consistency of the precision matrix estimate in variance estimation in two widely used portfolios in practice. The first one is Global Minimum Variance Portfolio (GMV) and the second one is Markowitz portfolio. Theorem 3.2 of \cite{caner2019} establishes consistency of these portfolio variance estimates at rate $\bar{s} \sqrt{logp/n}$, which is the rate of convergence of the precision matrix. This shows why precision matrix is the key in investment decisions.




In sum, naive nodewise regression is model-free, sparse and consistent. Sparsity assumption on precision matrix of outcomes can be thought of as a strong assumption.

\subsection{Residual Nodewise Regression}

This is an extension of the naive nodewise regression. It combines observed factors with nodewise regression. The main idea here is to use nodewise regression on precision matrix of errors and then by help of SMW formula, obtain precision matrix for asset returns (excess over T-Bill) through factor model. There is a major roadblock since nodewise regression uses feasible-observable random variables. Unobserved precision matrix of errors and estimating them is quite a challenge. \cite{caner2022} propose to run OLS, the outcome variables on factors and then using these residuals build a precision matrix estimate for errors. They show that this is a consistent way of estimating. Then using SMW formula, they obtain precision matrix  and its estimator for asset returns. They apply the precision matrix estimate to derive Sharpe Ratio estimates in several portfolios and show high dimensional consistency.

\subsubsection{Model}

The excess asset returns at time t for an asset j is represented as $y_{jt}$ and described by the following model
\[ y_{jt}= b_j' f_t + u_{jt},\]
where $j=1,\cdots,p$ represents assets, and $t=1,\cdots,n$ represents time span of the portfolio. We define the factor loadings $b_j: K \times 1$ vector, and $f_t: K \times 1$ vector of time varying observed factors, and $u_t$ is the unobserved errors.  In vector form
\begin{equation}
y_j = X' b_j + u_j,\label{rnw-1}
\end{equation}
where $X=(f_1, \cdots, f_n): K \times n$ matrix, and $y_j: n \times 1$ vector. Define the covariance matrix of errors as 
\[ \Sigma_u:= E (u_t u_t') : p \times p,\]
with $u_t= (u_{1t}, \cdots, u_{jt}, \cdots, u_{pt})'$. Then define the precision matrix of errors as $\Omega:= \Sigma_u^{-1}$. We impose sparsity 
on the rows of $\Omega$ and define 
\[ S_j=\{ j: \Omega_{jl} \neq 0 \},\]
and cardinality of nonzero cells in jth row of $\Omega$ as $|S_j|:= s_j$ and define the maximum number of nonzero cells across rows
as $\bar{s}:= \max_{1 \le j \le p} s_j$, and note that $\bar{s}$ is nondecreasing in $n$.

In the same way as in previous section, the main building block for precision matrix estimation is the following lasso regression
\begin{equation}
\tilde{\gamma}_j:= argmin_{\gamma_j \in R^{p-1}} \{ \| u_j - U_{-j}' \gamma_j \|_n^2 + 2 \lambda_n \| \gamma_j \|_1\},\label{rnw-2}
\end{equation}
where $U_{-j}': n \times p-1 $ matrix of errors except $j$ the asset error. $u_j: n \times 1$ vector is the error term for $j$ th asset across time. $\lambda_n>0$ is a tuning parameter, that is positive. Next applying (\ref{nw1})(\ref{nw2}) but with $\Sigma_u$ here we obtain the following linear regression
\begin{equation}
u_j = U_{-j}' \gamma + \eta_j,\label{rnw2a}
\end{equation}
and clearly this is an infeasible regression unlike previous section where we regress $y_j$ on $Y_{-j}$. This is one of the main differences with naive nodewise regression.  So \cite{caner2022} starts using OLS residuals instead of the errors. Namely, we start with 
\begin{equation}
\hat{u}_j= y_j - X' \hat{b}_j = M_X u_j,\label{rnw-3}
\end{equation}
where we use $\hat{b}_j= (X X')^{-1} X y_j$, and $M_X= I_n - X' (X X')^{-1} X$ and (\ref{rnw-1}) for the last equality in (\ref{rnw-3}).

Next, we need the residual matrix for the errors
\[ U_{-j} = Y_{-j} -  B_{-j} X,\]
where $Y_{-j}:  p-1 \times n $ matrix of excess asset returns except asset $j$, and $B_{-j}: p-1 \times K$ represents the factor loadings with $j$ th asset excluded. The residual matrix with $\hat{B}_{-j}$ as the OLS factor loadings estimate is
\begin{eqnarray}
 \hat{U}_{_j}' &= &Y_{-j}' - X' \hat{B}_{-j}' \nonumber \\
 & = & U_{-j}' - X' (X X')^{-1} X U_{-j}' \nonumber \\
 & = & M_X U_{-j}',\label{rnw-4}
 \end{eqnarray}
 and $\hat{U}_{-j}': n \times p-1$ matrix of residuals (transposed), and with $\hat{B}_{-j}'= (X X')^{-1} X Y_{-j}'$. So by premultiplying (\ref{rnw2a}) by $M_X$, we get (\ref{rnw-4})
 \begin{equation}
 \hat{u}_j = \hat{U}_{-j}' \gamma_j + \eta_{xj},\label{rnw-5}
 \end{equation}
with $\eta_{xj}= M_X \eta_j$. So we see from (\ref{rnw-5}) (\ref{rnw-2}) that $\gamma_j$ can be estimated with residuals as well. We now define a feasible nodewise estimator for off-diagonal terms in the precision matrix of errors. We start with the following 
\begin{equation}
\hat{\gamma}_j:= argmin_{\gamma_j \in R^{p-1}} \{ \| \hat{u}_j - \hat{U}_{-j}' \gamma_j \|_n^2 + 2 \lambda_n \| \gamma_j \|_1\}.\label{rnw5a}
\end{equation}
Then to setup the estimator for each row of the precision matrix for errors, take the $j$ th row and the main diagonal element estimator is
\[ \hat{\Omega}_{j,j}= \frac{1}{\hat{\tau}_j^2},\]
with $\hat{\tau}_j^2= \hat{u}_j' (\hat{u}_j - \hat{U}_{-j}' \hat{\gamma}_j)/n$, and the off-diagonal terms in $j$ th row are estimated by 
\[ \hat{\Omega}_{j,-j}= \frac{-\hat{\gamma}_{-j}'}{\hat{\tau}_j^2}.\]
We can write the $j$ th row of $\hat{\Omega}$ as 
\[ \hat{\Omega}_j'= \frac{\hat{C}_j'}{\hat{\tau}_j^2},\]
with $\hat{C}_j'$ being an $1 \times p$ vector with $j$ th element as 1, and the others are $-\hat{\gamma}_j'$. Then stacking all rows of $\hat{\Omega}_j'$ with $j=1,\cdots,p$ we obtain $\hat{\Omega}$.

We provide the assumptions before our precision matrix estimator for errors. Define $u_t$ as the $p \times 1$ error vector, and $u_{jt}$ is the $j$ the cell in that vector, and $U_{-jt}: p-1 \times 1$ vector of errors without $j$ th error at time $t$. Define the term $\eta_{jt}:= u_{jt} - U_{-jt}' \gamma_j$.

\begin{assum}\label{rnwa1}

(i). $u_t, f_t$ are sequences of strictly stationary and ergodic random variables. Furthermore $u_t$ and $f_t$ are independent. $u_t$ is $p \times 1$ mean zero random vector with covariance matrix $\Sigma_u$. $Eigmin (\Sigma_u)  \ge c > 0$, with $c$ a positive constant, and $\max_{1 \le j \le p} E u_{jt}^2 \le C < \infty$. (ii). For the strong mixing variables $f_t, u_t$, $\alpha(t) \le exp (-C t^{r_0})$ for a positive constant $r_0>0$.

\end{assum}

\begin{assum}\label{rnwa2}

There exist positive constants $r_1, r_2, r_3 >0$ and another set of positive constants $B_1, b_2, b_3, s_1, s_2, s_3 >0$ and for $t=1,\cdots,n$ and $j=1,\cdots,p$ with $k=1,\cdots, K$

(i). \[ P [ |u_{jt} | > s_1] \le exp [-(s_1/B_1)^{r_1}].\]

(ii). \[ P [ | \eta_{jt} | > s_2 ] \le exp [ - (s_2/b_2)^{r_2}].\]

(iii). \[ P [ | f_{kt} | > s_3] \le exp [ -(s_3/b_3)^{r_3}].\]

(iv). There exists $0 < \gamma_1 <1$ such that $\gamma_1^{-1}= 3 r_1^{-1} + r_0^{-1}$ and we also assume $3 r_2^{-1} + r_0^{-1} >1$ and 
$ 3 r_3^{-1} + r_0^{-1}>1$.

\end{assum}

Define $\gamma_2^{-1}:= 1.5 r_1^{-1}+ 1.5 r_2^{-1}+ r_0^{-1}$, and $\gamma_3^{-1}:= 1.5 r_1^{-1} + 1.5 r_3^{-1} + r_0^{-1}$ and let 
$\gamma_{min}:= min (\gamma_1, \gamma_2, \gamma_3)$. 

\begin{assum}\label{rnwa3}

$ [ln p]^{(2/\gamma_{min})-1} =o(n),$ and (ii). $K^2 = o(n)$, (iii). $K =o(p)$.

\end{assum}

\begin{assum}\label{rnwa4}
(i). $Eigmin (\Sigma_f) \ge c > 0$ with $\Sigma_f$ being the covariance of factors $f_t$, $t=1,\cdots,n$ (ii). $\max_{1 \le k \le K} E (f_{kt}^2) \le C < \infty$, 
$\min_{1 \le k \le K} E (f_{kt}^2) \ge c> 0$. (iii). $\max_{1 \le j \le p} E (\eta_{jt}^2) \le C < \infty$.

\end{assum}

\begin{assum}\label{rnwa5}
$\bar{s}, K, p$, and $n$ are such that (i). $ K^2 \bar{s}^{3/2} ln p/n \to 0$. (ii). $\bar{s} \sqrt{lnp/n} \to 0$.

\end{assum}

Assumptions \ref{rnwa1}-\ref{rnwa3} are standard and also used in \cite{fan2011}. By Assumption \ref{rnwa3}
$\sqrt{lnp/n}=o(1)$. Stationary GARCH with finite second moments with continuous error distributions, certain class of stationary Markov Chains, causal ARMA processes satisfy Assumptions \ref{rnwa1}, \ref{rnwa2}. Assumption \ref{rnwa4} is standard in linear factor models. Assumption \ref{rnwa5} is a sparsity assumption. This basically restricts the number of factors and number of nonzero cells in precision matrix, but allows $p>n$. The price of large $p$ is just square root log p, which is a small price to pay.

Now we can formally define the tuning parameter in penalization in lasso-nodewise regression which is in (\ref{rnw5a})
\begin{equation}
\lambda_n:= C [ \bar{K}^2 \bar{s}^{1/2} \frac{ln p}{n} + \frac{\sqrt{ln p}}{\sqrt{n}}].\label{rnw6}
\end{equation}
The second right side term in (\ref{rnw6}) is what we observe as the rate in  studies without factors (see  \cite{caner2019}) in high dimensional context. However, due to using residuals, and since the residuals  depend on growing number of factors, they affect the rate in tuning parameter. We need larger tuning parameter here compared to cross section studies. Under the Assumptions \ref{rnwa1}-\ref{rnwa5}, the high dimensional consistency for precision matrix of errors is achieved in Theorem 1 of \cite{caner2022}.
The rate of convergence is $\bar{s} \lambda_n$.

Next we connect the precision matrix of errors to precision matrix of excess asset returns. This will be done through a matrix inversion formula that is called Sherman-Morrison-Woodbury (SMW) formula as can be seen in \cite{hj2013}. It is described here. But before that note that covariance matrix of excess asset returns can be expressed as 
\[ \Sigma = B \Sigma_f B' + \Sigma_u,\]
where $B: p \times K$ factor loading matrix, and can be written as its $j$ th row is $b_j': 1 \times K $, and all the remainder of the matrix is $B_{-j}: p-1 \times K$. 
Setting $\Omega:= \Sigma_u^{-1}$ and the residual nodewise regression estimate for $\Omega$ as $\hat{\Omega}$, through SMW formula
, by defining the precision matrix as $\Gamma:= \Sigma^{-1}$ 
\[ \Gamma = \Omega - \Omega B [ \Sigma_f^{-1} + B' \Omega B]^{-1} B' \Omega,\]
with its estimate given by 
\begin{equation}
\hat{\Gamma}= \hat{\Omega} - \hat{\Omega} \hat{B} [ (\hat{\Sigma}_f)^{-1} + \hat{B}' \hat{\Omega}_{sym} \hat{B}]^{-1} \hat{B}' \hat{\Omega},\label{rnw7}
\end{equation}
with $\hat{\Omega}_{sym}:= \frac{\hat{\Omega}+ \hat{\Omega}'}{2}$ which is the symmetric version of $\hat{\Omega}$, and $\hat{\Sigma}_f= 
n^{-1} X X' - n^{-2} X 1_n 1_n' X'$ as the covariance matrix of returns, with $1_n$ representing a $n \times 1$ vector of ones. We also denote 
$\hat{B}= (Y X') (X X')^{-1}$ which is OLS based estimator.

We need two more assumptions on factor loadings and noise to get consistency for precision matrix estimate for outcomes.

\begin{assum}\label{rnwa6}
The factor loadings are such that 

(i).\[  \max_{1 \le j \le p} \max_{ 1 \le k \le K} | b_{jk}| \le C < \infty\]

(ii). \[ \| p^{-1} B'B - \Delta \|_{l_2} = o(1),\]
for some $K \times K$ symmetric positive definite matrix $\Delta$ such that $Eigmin (\Delta)$ is bounded away from zero.

\end{assum}

We need strengthened sparsity assumption for precision matrix of excess asset returns compared to errors in Assumption \ref{rnwa5}.

\begin{assum}\label{rnwa7}
Assume that 

(i). $Eigmax (\Sigma_u) \le C r_n$ with $C >0$ a positive constant, and $r_n \to \infty$ as $n \to \infty$, with $r_n/p \to 0$, and $r_n$ is a positive sequence.

(ii). $\bar{s} l_n \to 0$, with 
\[ l_n:= r_n^2 K^{5/2} max \left( \bar{s} \lambda_n, \bar{s}^{1/2} K^{1/2} \frac{\sqrt{max (ln p, ln n)}}{\sqrt{n}}\right).\]

\end{assum}

Clearly Assumption \ref{rnwa6} allows for pervasive factors which is a strong assumption. Assumption \ref{rnwa7} makes a realistic unbounded noise assumption, with larger $n$. Then the sparsity assumption here is a bit restrictive at least compared to precision matrix of errors one in Assumption \ref{rnwa5}, and also restrictive compared to nodewise sparsity condition in Assumption \ref{rnwa4}. This teaches us that there is an advantage with nodewise versus residual nodewise, in terms of sparsity. Number of factors play a negative role too. But again the consistency result is established under $p>n$ in Theorem 2 of \cite{caner2022} under Assumptions \ref{rnwa1}-\ref{rnwa4}, \ref{rnwa6}-\ref{rnwa7} again as in  \cite{caner2019}. We see that unlike nodewise regression due to growing number of factors, we need a stronger sparsity condition as in Assumption \ref{rnwa7}(ii). The rate of convergence is much slower at $\bar{s} l_n$. Also \cite{caner2022} obtains rate of convergence of Global Minimum Variance (GMV) portfolio Sharpe Ratio estimator at rate of 
$K^{3/2} \bar{s} l_n$, and this clearly shows that Sharpe Ratio estimation is difficult due to growing number of factors compared to only precision matrix estimation. Markowitz portfolio Sharpe Ratio  is also estimated in \cite{caner2022} but at rate $K^3 \bar{s} l_n$, which shows difficulty of estimating this compared to GMV. This is due to $\mu$ being used and involves number of factors compared to GMV, which there is no $\mu$, the expected return of the portfolio.

Next we provide an algorithm and then discuss this method's advantages and disadvantages.

\subsubsection{Algorithm to run Residual Nodewise Regression}

\begin{enumerate}

\item First run OLS  and get factor loading estimates for  $j$ th asset return, $y_j$ on $X$ and get $\hat{b}_j= (X X')^{-1} X y_j: K \times 1$.

\item Get a residual $\hat{u}_j= y_j - X' \hat{b}_j$. $X=(f_1, \cdots, f_n): K \times n$ matrix of factors. 

\item Form the transpose matrix of residuals for all asset returns except $j$ th one $\hat{U}_{-j}': n \times p-1$
\[ \hat{U}_{-j}' = Y_{-j}' - X' \hat{B}_{-j}',\]
with $\hat{B}_{-j}'= (X X')^{-1} X Y_{-j}': K \times p-1$ which is transpose of estimated factor loadings, and $Y_{-j}': n \times p-1$ matrix of asset returns except $j$ th one.

\item Run lasso $\hat{u}_j$ on $\hat{U}_{-j}'$, obtain $\hat{\gamma}_j$, and for the fixed $\lambda$ in that lasso regression, obtain $SSR (\lambda)$ as the sum of squares from the lasso regression $\hat{u}_j $ on $\hat{U}_{-j}'$, and $q(\lambda)$ represents the number of  nonzero cells lasso estimate $\hat{\gamma}_j$.

\item Obtain Generalized Information Criterion for a fixed $\lambda$ in a grid of $\Lambda=(\lambda_1,\cdots, \lambda_h)$ for $h $ possible $\lambda$ values
 \[ GIC (\lambda)= \frac{SSR (\lambda)}{n} + q(\lambda) log (p-1) \frac{ln (ln n)}{n},\]
and choose the minimum one, and pair with $\hat{\gamma}_j$   that corresponds to optimal $\lambda$. For each $j$, use the same optimal $\lambda$ choice., 

\item Then to get $\hat{\tau}_j^2$, use
\[ \hat{\tau}_j^2:= \hat{u}_j' (\hat{u}_j - \hat{U}_{-j}'. \hat{\gamma}_j)/n.\]

\item Form $\hat{\Omega}_j'$, which is $j$ th row of of precision matrix of errors, and the main diagonal term is $1/\hat{\tau}_j^2$, and the off-diagonal terms are 
$-\hat{\gamma}_j'/\hat{\tau}_j^2$.

\item Repeat steps 1-7 for all $j=1,\cdots,p$. Stack all rows $j=1,\cdots,p$ to form the precision matrix for errors $\hat{\Omega}$. Form symmetric version by $\hat{\Omega}_{sym}:= \frac{\hat{\Omega}+ \hat{\Omega}'}{2}$. 

\item Form 
\[ \hat{B}= (Y X') (XX')^{-1},\]
where $Y: p \times n$ matrix of asset returns, with $j=1,\cdots,p$ rows represent asset returns, and $t=1,\cdots, n$ is the time period. 

\item Form the covariance matrix estimate of factors
\[ \hat{\Sigma}_f := n^{-1} X X' - n^{-2} X 1_n 1_n' X'.\]

\item Form the precision matrix of estimates via (\ref{rnw7}) via steps 8-10.

\end{enumerate}

We now compare residual nodewise regression estimator with others. First of all, this is a model based estimator, but it is dense, and high dimensional consistent.
Compared to nodewise regression main advantage is, not assuming sparsity on asset return precision matrix, a disadvantage is, you have to estimate OLS based factors. If there is a mistake, that will carry over to the precision matrix estimation. Both of them are consistent in high dimensions. Compared to factor model based estimators such as \cite{fan2011}, the main difference is that, instead of thresholding the covariance of errors, a residual nodewise regression based precision matrix of errors are estimated. This is an advantage of residual nodewise, since it allows dense covariance matrix of errors. Compared to shrinkage based models, this is high dimensional consistent, but it is model based.

\section{Objective Functions}

One of the key discoveries in empirical practice in Magnus AI (a fin-tech startup in Izmir Turkey that uses techniques and objectives here), having a joint analysis of techniques above with objective functions. The main insight will be investing with the winner of joint technique and objective function. Separate analysis of technique from objectives is not rewarded. This analysis will be seen done in empirical section. The reason that a technique paired with an objective function, not joint analysis, can bring lower returns or Sharpe  Ratio. We may be tempted to use Nonlinear Shrinkage with  Global Minimum Variance Portfolio, but it may be the case that the winner in actual out-of-sample exercise may be Nodewise regression with Maximum Sharpe ratio portfolio. 

On the same issue, a lot of times investors are fixated on maximum Sharpe Ratio portfolio, thinking this may reward them. Since it has the best risk-return relation. However, a small risk with moderate return can give you a large  Sharpe Ratio. But there may be another portfolio that may reward you with higher risk but much higher return, with a low Sharpe ratio.

We start introducing three objective functions that are heavily used in practice and can earn returns. The first one is Global Minimum Variance (GMV) portfolio, and the second one is Markowitz-mean-variance portfolio (MW), and the last one is Maximum out-of-sample Sharpe Ratio (MSR) portfolio.

\subsection{GMV portfolio}

First we explain the GMV. The weights are given by 
\[ w_u^*:= argmin_{w \in R^p} [(w' \Sigma w) \quad  {\mbox s.to} \quad w'1_p=1],\]
where $1_p$ is a $p \times 1$ vector of ones.  As before, $\Sigma$ is the covariance matrix of returns. Variance of the portfolio is 
$w' \Sigma w$, and all weights add up to one, is a constraint that is helpful in communicating weights in the portfolio to customers. Define $\Theta:= \Sigma^{-1}$.
The solution weight vector is, 
\[ w_u^*= \frac{\Theta 1_p}{1_p' \Theta  1_p},\]
and this clearly shows the key aspect of precision matrix of asset outcomes $\Theta: p \times p$. For example, in a portfolio of $p=600$ stocks-assets, it will be very difficult to estimate with $n=90$, a train period of 90 days. 

So we minimize the variance subject to all weights are added to one. Solving the GMV and substituting this solution in variance we have 
\begin{equation}
 w_u^{*'} \Sigma w_u^* = [ 1_p' \Theta 1_p/p]^{-1}.\label{gmv1}
 \end{equation}
We can take into account large $p$ and for the proofs in papers by \cite{caner2019} and  \cite{caner2022} the scaling above is used to stabilize the variance.

Portfolio return will be $w_u^{*'} \mu$, in which $\mu:= E y_t: p \times 1 $ expected return vector for assets. Remember that $y_t$ represent excess asset return of $p$ assets.
Hence we can write that 
\[ w_u^{*'} \mu = \frac{1_p' \Theta \mu/p}{1_p' \Theta1_p/p},\]

Sharpe ratio of this portfolio is the return-risk ratio which is 
\[ SR_{GMV}:= \frac{w_u^{*'} \mu}{\sqrt{w_u^{*'} \Sigma w_u^*}} = \left( 1_p' \Theta \mu/p \right)
\left(1_p' \Theta 1_p/p
\right)^{-1/2}.\]

 Let  $\hat{\Theta}$ denote the precision matrix estimate in any of the methods, and $\hat{\mu}:=n^{-1} \sum_{t=1}^n y_t$ denoting the sample average of excess asset returns across time.  The variance estimate of the portfolio is 
 \[ \widehat{w' \Sigma w}= (1_p' \hat{\Theta} 1_p/p)^{-1}.\]
 Note that the variance estimator is the estimator for the expression on the right side of (\ref{gmv1})
 Then the estimate for the portfolio return is
 \[ \hat{w}' \hat{\mu}= \frac{1_p' \hat{\Theta} \hat{\mu}/p}{1_p' \hat{\Theta} 1_p/p}.
 \]

 It is very easy to see the importance of precision matrix in returns-variance and Sharpe Ratio of the portfolio.  This becomes key in large dimensional portfolios especially. To make things scaleable in large $p \to \infty$ we can do the following changes to estimator of Sharpe Ratio
\ \[ \widehat{SR}_{GMV}=  (1_p' \hat{\Theta} \hat{\mu}/p)(1_p' \hat{\Theta} 1_p/p)^{-1/2}.\]

Note that among the techniques that are discussed, nodewise regression of \cite{caner2019} analyzes the variance of GMV portfolio and obtains high dimensional consistency and residual nodewise regression of \cite{caner2022} obtains high dimensional consistency of GMV Sharpe ratio. These two techniques also deliver rates of convergence, and these are very briefly discussed in precision matrix estimation Subsections 5.1 and 5.2 respectively.

\subsection{Markowitz Portfolio}

Note that GMV portfolio is very risk averse. It involves only the variance and not the return. It is mainly used to develop portfolios, consisting of other portfolios, and intend to be a good device to provide a small return but not carry too much risk. A different portfolio for less risk averse investors can be Markowitz. It maximizes portfolio return for given risk, or its dual, minimize risk subject to a return. By choosing larger return-targets, a different risk is minimized.

  In the next portfolio, 
\[ w_m^*:= argmin_{w \in R^p} (w' \Sigma w), \quad {\mbox s.to} \quad w'1_p=1, \quad w'\mu=\rho_1,\]
where $\mu:= E y_t$, which is the expected asset returns, and $\rho_1$ is set as the expected portfolio return. $\rho_1$ will be given and $\mu$ will be estimated by sample average of returns
$\hat{\mu}:= n^{-1} \sum_{t=1}^n y_t$. 
The optimization problem can be solved and with $\Theta:= \Sigma^{-1}$ 
\[ w_m^*:= \left[ \frac{D_1 - \rho_1 F_1}{A_1D_1 - F_1^2}
\right] (\Theta  1_p) + \left[  \frac{\rho_1 A_1 - F_1}{A_1D_1 - F_1^2} \right] (\Theta \mu ) 
,\]
where we use $A_1:= 1_p' \Theta 1_p$, $F_1:= 1_p' \Theta \mu$ $D_1:=\mu' \Theta \mu $.

Define the scaled versions of above terms $A:=A_1/p, D:=D_1/p, F:=F_1/p$.
 The optimal portfolio variance will be given by 
\[ w_m^{*'} \Sigma w_m^*=  \frac{1}{p}\left[ \frac{A \rho_1^2 - 2 F \rho_1 + D]}{A D - F^2}
\right]\]

The portfolio return is:
\[ w_m^{*'} \mu = \rho_1,\]
through the constraint.

Sharpe Ratio of Markowitz portfolio is:
\[ SR_{MW}= \frac{w_m^{*'} \mu }{\sqrt{w_{m}^{*'} \Sigma_y w_{m}^*}}= \rho_1 \frac{p^{1/2} \sqrt{ AD - F^2}}{\sqrt{A \rho_1^2 - 2 F \rho_1 + D}}.\]

It is clear that only the variance and Sharpe Ratio estimation is needed. The variance estimator is, with $\hat{A}:= 1_p' \hat{\Theta} 1_p/p$, $\hat{F}:= 1_p' \hat{\Theta} \hat{\mu}/p$,
$\hat{D}= \hat{\mu}' \hat{\Theta} \hat{\mu}/p$.

\[ \widehat{w_m^{*'} \Sigma_y w_m^*}=  \frac{1}{p} \left[ \frac{\hat{A} \rho_1^2 - 2 \hat{F} \rho_1 + \hat{D}}{\hat{A}\hat{ D} - \hat{F}^2}
\right]\]

The estimator for Sharpe Ratio is:
\[ \widehat{SR}_{MW}:= \rho_1  \frac{p^{1/2} \sqrt{ \hat{A}\hat{D} - \hat{F}^2}}{\sqrt{\hat{A} \rho_1^2 - 2 \hat{F} \rho_1 + \hat{D}}}\]

Note that again nodewise regression paper of \cite{caner2019} estimates the variance consistently in high dimensions, and residual nodewise of \cite{caner2022} estimates the Sharpe Ratio consistently in high dimensions. Rates of convergences are derived and discussed in Sections 5.1, 5.2.

\subsection{Maximum Out-of-Sample Sharpe Ratio Portfolio}

We proceed with maximum Sharpe Ratio portfolio. This is used heavily in practice and academic research since it is a simple and elegant metric with a closed form solution.
It tradeoffs risk and return. It is similar to Markowitz portfolio above but finds the best tradeoff rather than using any given tradeoff. 
\[ w_{MSR}:= argmax_{w \in R^p} [w' \mu \quad {\mbox s.to} \quad w' \Sigma w \le \sigma^2],\]
for a given known $\mu,\sigma^2$. 
The solution is with $\Theta:=\Sigma^{-1}$,
\[ w_{MSR}:=  \frac{\sigma \Theta \mu}{\sqrt{\mu' \Theta \mu}}.\]

 The variance for the portfolio simplifies, and  
\[ w_{MSR}' \Sigma w_{MSR}= \sigma^2.\] 
The portfolio return is
\[ w_{MSR}' \mu = \sigma \sqrt{ \mu' \Theta \mu/p} .\] 
The Sharpe Ratio is:
\[ SR_{MSR}= \frac{w_{MSR}' \mu}{\sqrt{w_{MSR}' \Sigma w_{MSR}}} = \sqrt{\mu' \Theta \mu/p}.\]

The estimators  will use $\hat{\Theta}$, and $\hat{\mu}$ as sample average of returns for $\mu$ estimation. 
But since this is an out-sample metric, we keep $\mu, \Sigma$ as given and only put estimates of weights

Hence
the estimator for out-sample variance is, with $\hat{\mu}$ as the sample average of returns only used for weight estimation, and $\hat{\Theta}$ is one of the techniques described for precision matrix estimation,
\[ (\widehat{w_{MSR}})' \Sigma (\widehat{w_{MSR}})= \sigma^2 \frac{(\hat{\mu} \hat{\Theta}/p)' \Sigma
(\hat{\mu} \hat{\Theta}/p)
  }{(\hat{\mu}' \hat{\Theta} \hat{\mu}/p )}.\] 
and  estimate for out-sample portfolio return is:
\[ (\widehat{w_{MSR}})' \mu = \sigma \frac{\hat{\mu}' \hat{\Theta}' \mu/p }{\sqrt{\hat{\mu}' \hat{\Theta} \hat{\mu}/p}}.\] 
and the Sharpe Ratio estimate is:
\[ \widehat{SR_{MSR}}= \frac{\hat{\mu}' \hat{\Theta}' \mu/p }{\sqrt{(\hat{\mu} \hat{\Theta}/p)' \Sigma 
(\hat{\mu} \hat{\Theta}/p)}
  }.\]

Note that all formulas heavily depend on $\hat{\Theta}$, hence it is crucial in portfolio analysis. \cite{caner2022} residual nodewise regression shows that Sharpe Ratio estimates are consistent and at a rate $K^2 \bar{s} l_n$ which is faster than Markowitz rate due to out-sample $\mu,\Sigma$ usage.

\section{Empirics}

This section considers performance of method-objective functions jointly as used by Magnus-AI startup in Turkey.

The stock returns and historical constituents of S\&P 500 stocks are from CRSP in the period covering 1985-01 to 2024-12.  The time-series data of Fama-French 5 factors and risk-free rates are from Kenneth R. French - Data Library. All data are in monthly frequency. The out of sample period is shown on the caption of each table. In computations, we use the no-looking-forward rolling window method as in \cite{bs2022}
with window size $T_I=180$, and use the transaction costs as in \cite{caner2022}, with the transaction costs set to 50 bps.  So the in-sample period is 180 months before from the beginning of out sample month. For example, for  TABLE 1, out-sample is 2000-01 (January) to 2003-12 (December), the in-sample period is 1985-01 to 1999-12. In that example our number of assets is 500, and the time span is 180.

We compare the performance of the following machine learning methods and include S\&P 500 index to compare as a benchmark.
\begin{itemize}
    \item 	NW: nodewise introduced by \cite{caner2019} in Section 5.
    \item 	Residual-based Nodewise of  \cite{caner2022} in Section 5.
    \item 	POET: Latent factor model of \cite{fan2013} in Section 4.
    \item 	OFT: POET with FF 5-factors as observed factors via \cite{fan2011}.
    \item 	LSLW: linear shrinkage of \cite{lw2004} in Section 3.
    \item 	NLS: nonlinear shrinkage of \cite{lw2017} in Section 3.
    \item 	SFNL: single factor nonlinear shrinkage of \cite{lw2017} in Section 3.
    \item 	S\&P 500: benchmark index.

\end{itemize}

We consider the following performance measures.

Since we only report the net (of transaction costs) returns related measures of the portfolios not subject to survival-ship bias, we first provide the transaction cost details (Net Return Adjustment) in the following:

\[y_{P,t+1}^{Net} = y_{P,t+1} - c (1 + y_{P,t+1}) \sum_{j=1}^p |\hat{w}_{t+1,j} - \hat{w}_{t,j}^+|,\]

$y_{P,t+1}^{Net} $ is the portfolio's net (of transaction costs) return in Out Of Sample $t+1$. $y_{P,t+1}  = \hat{w}_t' y_{t+1}$ is the gross portfolio return at $t+1$, with $\hat{w}_t$ representing the estimated weight of the portfolio at time $t$, with one of the methods. Set
$c = 0.005$ (50 basis points, following \cite{caner2022}). Then, with $\hat{w}_{t,j}$ representing the estimated weight of asset $j$ at time $t$,
$\hat{w}_{t,j}^+ = \hat{w}_{t,j} \frac{1 + y_{t+1,j}}{1 + y_{P,t+1}}$: Adjusted weights at end of period $t$ ($y_{t,j}$ is the gross excess return of asset $j$ at time $t$). The above formula adjusts gross returns for proportional transaction costs on rebalancing. $y_{P,t+1}^{Net} $ is then used to compute net versions of return, variance, and Sharpe Ratio in the following. 

Out-of-Sample Average Return (with Transaction Costs) is defined as

\[ \hat{\mu}_{os}^{Net} = \frac{1}{T - T_I} \sum_{t=T_I}^{T-1} y_{P,t+1}^{Net}.\]
$y_{P,t+1}^{Net} $ is the portfolio's net (transaction costs) return in OOS $t+1$ defined above. $\hat{\mu}_{os}^{Net}$ is the average OOS portfolio return over rolling windows, reported as ``Return" in Tables.

Out-of-Sample Variance (with Transaction Costs) is defined as 

\[ \hat{\Sigma}_{NET,os}^2 = \frac{1}{T - T_I - 1} \sum_{t=T_I}^{T-1} (y_{P,t+1}^{Net} - \hat{\mu}_{os}^{Net})^2.\]
The above sample variance of OOS portfolio returns is reported as ``Variance" in the tables.

The Sharpe Ratio (SR) is calculated from the average net return and the variance of the net-of-cost portfolio returns in the out-of-sample period.

\[SR = \frac{\hat{\mu}_{os}^{NET}}{\hat{\Sigma}_{NET,os}}.\]

Turnover (rebalancing frequency, averaged over OOS windows) is defined as:
\[turnover = \sum_{j=1}^p |\hat{w}_{t+1,j} - \hat{w}_{t,j}^+|.\]

In the following tables, we output the performance measures combined with the different portfolios for different machine learning methods of a given out-of-sample period in one table. For example, for Sharpe ratio-SR, we report the Markowitz mean-variance portfolio: MV, Global Minimum Variance portfolio: GMV, Maximum Sharpe Ratio portfolio: MSR portfolios in one table, and their SR are MV-SR, GMV-SR, MSR-SR, respectively. The same format applies to Return, Variance and Turnover.
Tables below cover 5 distinct time periods. The first 3 time periods are chosen to see how ML-AI methods behave in stock market downturns, 2000/01-2003/12, 2008/01-2010/12, 2020/01-2023/12. They reflect dot-com bubble, financial recession, and Covid and 2022 stock market downturn respectively. There are also up moments in these periods but mainly, stock market underperformed.
It would be interesting since there is presumption that ML-AI are momentum techniques. Then we take a longer time perspective and consider the first decade of 2000/01-2010/12, and then increase the span in 21 st century until Covid, 2000/01-2019/12. There are other time periods analyzed but these 5 gave us short term downturns and ML-AI and long term with downturns included. The other time periods that we analyze are about mainly stock market booms 2010-2019, 2005-2019, and gave similar responses, but these are available on demand. At each table the winner is given by {\bf bold} faced number, and the runner up is given by {\it italics}. We also consider no short selling allowed but it did not change any of the findings with short selling, so we decided not to include, but these are also available from authors on demand.


\subsection{Sharpe Ratios}

The performance analysis of the portfolio optimization methods (NW, Residual-based NW, POET, OFT, LSLW, NLS, SFNL) compared to the S\&P 500 benchmark reveals distinct patterns across the evaluated time periods. During the early 2000s (2000/01-2003/12), a period marked by the dot-com bubble burst, most methods exhibit negative Sharpe Ratios in Table 1, particularly in the MSR portfolio. For instance, in Table 1 (SR: 2000/01-2003/12), Sharpe Ratios range from -1.2735 (OFT) to -0.0737 (NW) for MSR,  indicating poor risk-adjusted returns. The NW method consistently shows the least negative or slightly positive Sharpe Ratios (e.g., 0.0171 for MV), suggesting a degree of robustness in this volatile period. Nodewise with Markowitz portfolio is clear winner here beating $S\%P$ 500 and delivering positive Sharpe ratio among all methods. The runner up is again Nodewise but with GMV portfolio.

The 2008/01-2010/12 period in Table 2, encompassing the global financial crisis, presents a similar trend of underperformance. Sharpe Ratios remain negative across all methods, with OFT (-1.0347 for MSR) and LSLW (-1.1182 for MSR) showing the poorest risk-adjusted returns. The winner is again Nodewise-GMV combination with -0.041, and $S\&P$ 500 is the runner up.

The more recent 2020/01-2023/12 period Table 3, potentially influenced by post-pandemic recovery, shows a mixed outcome. The S\&P 500 achieves a positive Sharpe Ratio (0.1394), while most methods remain negative, except NW (0.0327 for GMV, 0.0280 for MSR).  The winner is $S\&P 500$  with  0.1394, and the runner up is Nodewise-GMV with 0.0327.

In Table 4, we see that in decade of 2000-2010, Nodewise-GMV has the best record with only positive Sharpe Ratio with 0.0007.
The runner up is again Nodewise with Markowitz Mean Variance portfolio -0.0029. Investing with $S\&P 500$ could have been a terrible idea, going along with index fund, since its SR is -0.0538. OFT with observed factor models also deliver very negative Sharpe Ratios between -0.6874/-0.7136.

Table 5 offers a longer perspective of 20 years just before Covid, 2000-2019. Except from Nodewise and $S\&P 500$ all ML methods deliver negative SRs. Nodewise-GMV is the long run winner with 0.0659 and the runner up is $S\%P500$ with 0.0580. 

Overall Nodewise-GMV delivered the best results, 3 times winner with 2 times runner up positions. It also delivered the best long term result in Table 5.

\subsection{Returns}

Tables 6-10 analyze returns. These are in line with SR-Tables 1-5. Nodewise-GMV is again 3 times winner and 2 times runner up in Tables 6-10. We analyze few tables since the results are the same as in Tables  1-5. It shows that returns are the main determinants of SR in those time periods. Having said that, in Table 6, 2000-2003 period, Nodewise Markowitz-mean variance portfolio was the winner with 0.0009, and the runner up is again Nodewise but with GMV. $S\%P500$ delivered negative returns with -0.0075. Tables 9-10 show that Nodewise GMV was the winner in 2000-2010, and 2000-2019 periods with zero return (approximated to fifth decimal)on average, and 0.0027 respectively. The other ML methods deliver negative returns in all Tables 6-10.

\subsection{Variance}
Tables 11-15 analyze variance.
Here clearly there are dominance of non-linear shrinkage models of \cite{lw2017}. Note that they are designed to minimize portfolio variance and they are very successful in that regard. Note that only once, residual nodewise is winner in Table 12 in period 2008-2010. There residual nodewise-MV gets 0.0436 as the variance and $S\%P 500$ has 0.0643 and one of the largest variances. In Table 15, in long term we see that Nonlinear Shrinkage-MV has the best variance at 0.0329, and the runner up is Single Factor-nonlinear shrinkage with 0.330. In all these Tables, we observe that passive investing with $S\%P 500$ brings large variance, and is one of the worst in Table 15 in the long run with 0.0421.

\subsection{Turnover}

Tables 16-20 consider turnover in ML based investing. Turnover is the best in $S\%P500$ of course, but among ML methods Nodewise-GMV is always the runner-up. To give an  example at Table 20, for 2000-2019 period Nodewise GMV has 0.0965 and SP500 has 0.0778.

\subsection{Summary of Results}

It is clear that 2008-2010 period is the most negative one in terms of obtaining good returns and Sharpe Ratios. In the other stock market downturn as in 2000-2003, and Covid and 2022 downturn: 2020-2023 we see that Nodewise with Markowitz-mean variance or GMV can deliver positive-non-negative Sharpe ratios. Long-run, 20 years of data,  as in 2000-2019 shows us again good performance of Nodewise-GMV. Note that passive indexing comes with a caveat, $S\%P500$ produces large variance, hence it is not very stable. Although Nodewise-GMV can produce large variance too, except from 2008-2010, its variance is lower than $S\&P500$.

\newpage

\begin{table}[H]
\begin{threeparttable}
\centering
\caption{SR: Out Sample:2000/01-2003/12\\ In-Sample 1985/01-1999/12. Short selling allowed.}
\begin{tabular}{lccc}
\toprule
Method & GMV-SR & MSR-SR & MV-SR \\
\midrule
NW & {\it 0.0000} & -0.0737 & {\bf 0.0171} \\
Residual-based NW & -0.3228 & -0.9043 & -0.3216 \\
POET & -0.3369 & -0.7281 & -0.3069 \\
OFT & -0.8495 & -1.2735 & -0.9511 \\
LSLW & -1.0882 & -1.2688 & -1.0711 \\
NLS & -0.7226 & -1.1029 & -0.7009 \\
SFNL & -0.6279 & -1.1592 & -0.7075 \\
S\&P 500 & -0.1453 & -0.1453 & -0.1453 \\
\bottomrule
\end{tabular}
\begin{tablenotes}
\small
\item  GMV=Global minimum variance portfolio, MV=Mean-variance portfolio with target returns = 1\% monthly. MSR=Maximum Sharpe ratio portfolio. Boldface number denotes the winner and italic one denotes the runner up.
\end{tablenotes}
\end{threeparttable}
\end{table}

\begin{table}[H]
\begin{threeparttable}
\centering
\caption{SR: Out Sample: 2008/01-2010/12\\In-Sample: 1993/01-2007/12. Short selling allowed.}
\begin{tabular}{lccc}
\toprule
Method & GMV-SR & MSR-SR & MV-SR \\
\midrule
NW & {\bf -0.0410} & -0.0881 & -0.0510 \\
Residual-based NW & -0.4473 & -0.5973 & -0.4448 \\
POET & -0.3893 & -0.5130 & -0.3836 \\
OFT & -0.7041 & -1.0347 & -0.7018 \\
LSLW & -0.7402 & -1.1182 & -0.7404 \\
NLS & -0.5500 & -0.9299 & -0.5467 \\
SFNL & -0.4829 & -0.9127 & -0.4877 \\
S\&P 500 & {\it -0.0503} & {\it -0.0503} & {\it -0.0503} \\
\bottomrule
\end{tabular}
\begin{tablenotes}
\small
\item GMV=Global minimum variance portfolio, MV=Mean-variance portfolio with target returns = 1\% monthly. MSR=Maximum Sharpe ratio portfolio.Boldface number denotes the winner and italic one denotes the runner up.
\end{tablenotes}
\end{threeparttable}
\end{table}

\newpage

\begin{table}[H]
\begin{threeparttable}
\centering
\caption{SR: Out-Sample: 2020/01-2023/12. \\
In-Sample: 2005/01-2019/12.Short selling allowed.}
\begin{tabular}{lccc}
\toprule
Method & GMV-SR & MSR-SR & MV-SR \\
\midrule
NW & {\it 0.0327} & 0.0280 & 0.0037 \\
Residual-based NW & -0.3881 & -0.6011 & -0.4013 \\
POET & -0.2724 & -0.4551 & -0.2978 \\
OFT & -0.9091 & -1.2812 & -0.8917 \\
LSLW & -0.8549 & -0.9403 & -0.8697 \\
NLS & -0.5257 & -0.6730 & -0.5394 \\
SFNL & -0.5220 & -0.8279 & -0.5017 \\
S\&P 500 & {\bf 0.1394} & {\bf 0.1394} & {\bf 0.1394} \\
\bottomrule
\end{tabular}
\begin{tablenotes}
\small
\item  GMV=Global minimum variance portfolio, MV=Mean-variance portfolio with target returns = 1\%. MSR=Maximum Sharpe ratio portfolio.Boldface number denotes the winner and italic one denotes the runner up.
\end{tablenotes}
\end{threeparttable}
\end{table}

\begin{table}[H]
\begin{threeparttable}
\centering
\caption{SR: Out-Sample: 2000/01-2010/12.\\
In-Sample: 1985/01-1999/12. Short selling allowed.}
\begin{tabular}{lccc}
\toprule
Method & GMV-SR & MSR-SR & MV-SR \\
\midrule
NW & {\bf 0.0007} & -0.0571 & {\it -0.0029} \\
Residual-based NW & -0.2677 & -0.6620 & -0.2650 \\
POET & -0.2710 & -0.5958 & -0.2552 \\
OFT & -0.6874 & -0.7142 & -0.7136 \\
LSLW & -0.8084 & -0.9861 & -0.8113 \\
NLS & -0.5076 & -0.8500 & -0.5027 \\
SFNL & -0.4677 & -0.7648 & -0.4906 \\
S\&P 500 & -0.0538 & -0.0538 & -0.0538 \\
\bottomrule
\end{tabular}
\begin{tablenotes}
\small
\item GMV=Global minimum variance portfolio, MV=Mean-variance portfolio with target returns = 1\%. MSR=Maximum Sharpe ratio portfolio.Boldface number denotes the winner and italic one denotes the runner up.
\end{tablenotes}
\end{threeparttable}
\end{table}

\begin{table}[H]
\begin{threeparttable}
\centering
\caption{SR: Out-Sample:2000/01-2019/12.\\
In-Sample: 1985/01-1999/12. Short selling allowed.}
\begin{tabular}{lccc}
\toprule
Method & GMV-SR & MSR-SR & MV-SR \\
\midrule
NW & {\bf 0.0659} & 0.0147 & 0.0525 \\
Residual-based NW & -0.1653 & -0.5444 & -0.1689 \\
POET & -0.0970 & -0.4527 & -0.0928 \\
OFT & -0.6829 & -0.5927 & -0.6939 \\
LSLW & -0.7559 & -0.8018 & -0.7602 \\
NLS & -0.4205 & -0.7088 & -0.4195 \\
SFNL & -0.4150 & -1.1430 & -0.4229 \\
S\&P 500 & {\it 0.0580} & {\it 0.0580} &  {\it 0.0580} \\
\bottomrule
\end{tabular}
\begin{tablenotes}
\small
\item GMV=Global minimum variance portfolio, MV=Mean-variance portfolio with target returns = 1\%. MSR=Maximum Sharpe ratio portfolio.Boldface number denotes the winner and italic one denotes the runner up.
\end{tablenotes}
\end{threeparttable}
\end{table}

\begin{table}[H]
\begin{threeparttable}
\centering
\caption{Return: Out-Sample: 2000/01-2003/12.\\
In-Sample: 1985/01-1999/12. Short selling allowed.}
\begin{tabular}{lccc}
\toprule
Method & GMV-Return & MSR-Return & MV-Return \\
\midrule
NW & {\it 0.0000} & -0.0030 & {\bf 0.0009} \\
Residual-based NW & -0.0156 & -0.1058 & -0.0154 \\
POET & -0.0156 & -0.0854 & -0.0142 \\
OFT & -0.0382 & -0.8016 & -0.0402 \\
LSLW & -0.0469 & -0.1979 & -0.0463 \\
NLS & -0.0259 & -0.1429 & -0.0253 \\
SFNL & -0.0239 & -0.3037 & -0.0253 \\
S\&P 500 & -0.0075 & -0.0075 & -0.0075 \\
\bottomrule
\end{tabular}
\begin{tablenotes}
\small
\item GMV:Global minimum variance portfolio, MV:Mean-variance portfolio with target returns = 1\%. MSR:Maximum Sharpe ratio portfolio.Boldface number denotes the winner and italic one denotes the runner up.
\end{tablenotes}
\end{threeparttable}
\end{table}

\begin{table}[H]
\begin{threeparttable}
\centering
\caption{Return: Out-Sample: 2008/01-2010/12. \\
In-Sample: 1993/01-2007/12.Short selling allowed.}
\begin{tabular}{lccc}
\toprule
Method & GMV-Return & MSR-Return & MV-Return \\
\midrule
NW &{\bf  -0.0028} & -0.0056 & -0.0034 \\
Residual-based NW & -0.0202 & -0.0492 & -0.0194 \\
POET & -0.0185 & -0.0471 & -0.0178 \\
OFT & -0.0358 & -0.1352 & -0.0352 \\
LSLW & -0.0376 & -0.1134 & -0.0375 \\
NLS & -0.0265 & -0.0871 & -0.0262 \\
SFNL & -0.0240 & -0.0967 & -0.0241 \\
S\&P 500 & {\it -0.0032} & {\it -0.0032} &  {\it -0.0032} \\
\bottomrule
\end{tabular}
\begin{tablenotes}
\small
\item GMV:Global minimum variance portfolio, MV:Mean-variance portfolio with target returns = 1\%. MSR:Maximum Sharpe ratio portfolio.Boldface number denotes the winner and italic one denotes the runner up.
\end{tablenotes}
\end{threeparttable}
\end{table}

\begin{table}[H]
\begin{threeparttable}
\centering
\caption{Return: Out-Sample: 2020/01-2023/12. \\
In-Sample: 2005/01-2019/12. Short selling allowed.}
\begin{tabular}{lccc}
\toprule
Method & GMV-Return & MSR-Return & MV-Return \\
\midrule
NW & {\it 0.0018} & 0.0015 & 0.0002 \\
Residual-based NW & -0.0163 & -0.0376 & -0.0169 \\
POET & -0.0128 & -0.0302 & -0.0139 \\
OFT & -0.0393 & -0.1056 & -0.0364 \\
LSLW & -0.0382 & -0.0732 & -0.0387 \\
NLS & -0.0208 & -0.0476 & -0.0213 \\
SFNL & -0.0217 & -0.0669 & -0.0203 \\
S\&P 500 & {\bf 0.0079} & {\bf 0.0079} & {\bf 0.0079} \\
\bottomrule
\end{tabular}
\begin{tablenotes}
\small
\item GMV=Global minimum variance portfolio, MV=Mean-variance portfolio with target returns = 1\%. MSR=Maximum Sharpe ratio portfolio.Boldface number denotes the winner and italic one denotes the runner up.
\end{tablenotes}
\end{threeparttable}
\end{table}

\begin{table}[H]
\begin{threeparttable}
\centering
\caption{Return: Out-Sample: 2000/01-2010/12. \\
In-Sample 1985/01-1999/12: Short selling allowed.}
\begin{tabular}{lccc}
\toprule
Method & GMV-Return & MSR-Return & MV-Return \\
\midrule
NW & {\bf 0.0000} & -0.0025 &  {\it -0.0001} \\
Residual-based NW & -0.0109 & -0.0611 & -0.0105 \\
POET & -0.0108 & -0.0543 & -0.0101 \\
OFT & -0.0283 & -0.3636 & -0.0287 \\
LSLW & -0.0336 & -0.1250 & -0.0336 \\
NLS & -0.0185 & -0.0893 & -0.0183 \\
SFNL & -0.0175 & -0.1556 & -0.0179 \\
S\&P 500 & -0.0026 & -0.0026 & -0.0026 \\
\bottomrule
\end{tabular}
\begin{tablenotes}
\small
\item GMV:Global minimum variance portfolio, MV:Mean-variance portfolio with target returns = 1\%. MSR:Maximum Sharpe ratio portfolio.Boldface number denotes the winner and italic one denotes the runner up.
\end{tablenotes}
\end{threeparttable}
\end{table}

\begin{table}[H]
\begin{threeparttable}
\centering
\caption{Return: Out-Sample: 2000/01-2019/12.\\
In-Sample: 1985/01-1999/12. Short selling allowed.}
\begin{tabular}{lccc}
\toprule
Method & GMV-Return & MSR-Return & MV-Return \\
\midrule
NW & {\bf 0.0027} & 0.0006 & 0.0022 \\
Residual-based NW & -0.0060 & -0.0409 & -0.0060 \\
POET & -0.0035 & -0.0337 & -0.0033 \\
OFT & -0.0243 & -0.2398 & -0.0243 \\
LSLW & -0.0283 & -0.0698 & -0.0283 \\
NLS & -0.0139 & -0.1147 & -0.0138 \\
SFNL & -0.0138 & -0.0645 & -0.0140 \\
S\&P 500 & {\it 0.0024} & {\it 0.0024} & {\it 0.0024} \\
\bottomrule
\end{tabular}
\begin{tablenotes}
\small
\item GMV=Global minimum variance portfolio, MV=Mean-variance portfolio with target returns = 1\%. MSR=Maximum Sharpe ratio portfolio.Boldface number denotes the winner and italic one denotes the runner up.
\end{tablenotes}
\end{threeparttable}
\end{table}

\begin{table}[H]
\begin{threeparttable}
\centering
\caption{Variance: Out-Sample:2000/01-2003/12.\\
In-Sample: 1985/0-1999/12. Short selling allowed.}
\begin{tabular}{lccc}
\toprule
Method & GMV-Variance & MSR-Variance & MV-Variance \\
\midrule
NW & 0.0469 & 0.0450 & 0.0509 \\
Residual-based NW & 0.0484 & 0.1170 & 0.0479 \\
POET & 0.0464 & 0.1173 & 0.0463 \\
OFT & 0.0450 & 0.6295 & 0.0422 \\
LSLW & 0.0431 & 0.1560 & 0.0433 \\
NLS & {\it 0.0359} & 0.1296 & 0.0361 \\
SFNL & 0.0380 & 0.2620 & {\bf 0.0357} \\
S\&P 500 & 0.0518 & 0.0518 & 0.0518 \\
\bottomrule
\end{tabular}
\begin{tablenotes}
\small
\item GMV=Global minimum variance portfolio, MV=Mean-variance portfolio with target returns = 1\%. MSR=Maximum Sharpe ratio portfolio.Boldface number denotes the winner and italic one denotes the runner up.
\end{tablenotes}
\end{threeparttable}
\end{table}

\begin{table}[H]
\begin{threeparttable}
\centering
\caption{Variance: Out-Sample: 2008/01-2010/12.\\
In-Sample: 1993/01-2007/12. Short selling allowed.}
\begin{tabular}{lccc}
\toprule
Method & GMV-Variance & MSR-Variance & MV-Variance \\
\midrule
NW & 0.0675 & 0.0639 & 0.0674 \\
Residual-based NW & {\it 0.0452} & 0.0824 & {\bf 0.0436} \\
POET & 0.0476 & 0.0919 & 0.0464 \\
OFT & 0.0508 & 0.1306 & 0.0501 \\
LSLW & 0.0508 & 0.1014 & 0.0507 \\
NLS & 0.0481 & 0.0936 & 0.0479 \\
SFNL & 0.0497 & 0.1059 & 0.0494 \\
S\&P 500 & 0.0643 & 0.0643 & 0.0643 \\
\bottomrule
\end{tabular}
\begin{tablenotes}
\small
\item GMV=Global minimum variance portfolio, MV=Mean-variance portfolio with target returns = 1\%. MSR=Maximum Sharpe ratio portfolio.Boldface number denotes the winner and italic one denotes the runner up.
\end{tablenotes}
\end{threeparttable}
\end{table}

\begin{table}[H]
\begin{threeparttable}
\centering
\caption{Variance: Out-Sample: 2020/01-2023/12.\\
In-Sample: 2005/01-2019/12. Short selling allowed.}
\begin{tabular}{lccc}
\toprule
Method & GMV-Variance & MSR-Variance & MV-Variance \\
\midrule
NW & 0.0558 & 0.0539 & 0.0560 \\
Residual-based NW & 0.0420 & 0.0626 & 0.0421 \\
POET & 0.0471 & 0.0663 & 0.0466 \\
OFT & 0.0433 & 0.0824 & 0.0408 \\
LSLW & 0.0446 & 0.0779 & 0.0445 \\
NLS & {\bf 0.0396} & 0.0708 & {\bf 0.0396} \\
SFNL & 0.0416 & 0.0809 & {\it 0.0404} \\
S\&P 500 & 0.0569 & 0.0569 & 0.0569 \\
\bottomrule
\end{tabular}
\begin{tablenotes}
\small
\item GMV=Global minimum variance portfolio, MV=Mean-variance portfolio with target returns = 1\%. MSR=Maximum Sharpe ratio portfolio.Boldface number denotes the winner and italic one denotes the runner up.
\end{tablenotes}
\end{threeparttable}
\end{table}

\begin{table}[H]
\begin{threeparttable}
\centering
\caption{Variance: Out-Sample:2000/01-2010/12.\\
In-Sample: 1985/01-1999/12. Short selling allowed.}
\begin{tabular}{lccc}
\toprule
Method & GMV-Variance & MSR-Variance & MV-Variance \\
\midrule
NW & 0.0465 & 0.0445 & 0.0479 \\
Residual-based NW & 0.0406 & 0.0922 & 0.0398 \\
POET & 0.0400 & 0.0911 & 0.0396 \\
OFT & 0.0411 & 0.5091 & 0.0402 \\
LSLW & 0.0416 & 0.1267 & 0.0414 \\
NLS & {\it 0.0365} & 0.1050 & {\bf 0.0364} \\
SFNL & 0.0373 & 0.2034 & 0.0366 \\
S\&P 500 & 0.0475 & 0.0475 & 0.0475 \\
\bottomrule
\end{tabular}
\begin{tablenotes}
\small
\item GMV=Global minimum variance portfolio, MV=Mean-variance portfolio with target returns = 1\%. MSR=Maximum Sharpe ratio portfolio.Boldface number denotes the winner and italic one denotes the runner up.
\end{tablenotes}
\end{threeparttable}
\end{table}

\begin{table}[H]
\begin{threeparttable}
\centering
\caption{Variance: Out-Sample: 2000/01-2019/12.\\
In-Sample: 1985/01-1999/12. Short selling allowed.}
\begin{tabular}{lccc}
\toprule
Method & GMV-Variance & MSR-Variance & MV-Variance \\
\midrule
NW & 0.0403 & 0.0392 & 0.0411 \\
Residual-based NW & 0.0363 & 0.0751 & 0.0357 \\
POET & 0.0361 & 0.0743 & 0.0357 \\
OFT & 0.0356 & 0.4026 & 0.0351 \\
LSLW & 0.0374 & 0.0871 & 0.0372 \\
NLS & 0.0331 & 0.1618 & {\bf 0.0329} \\
SFNL & 0.0333 & 0.0546 & {\it 0.0330} \\
S\&P 500 & 0.0421 & 0.0421 & 0.0421 \\
\bottomrule
\end{tabular}
\begin{tablenotes}
\small
\item GMV=Global minimum variance portfolio, MV=Mean-variance portfolio with target returns = 1\%. MSR=Maximum Sharpe ratio portfolio.Boldface number denotes the winner and italic one denotes the runner up.
\end{tablenotes}
\end{threeparttable}
\end{table}

\begin{table}[H]
\begin{threeparttable}
\centering
\caption{Turnover: Out-Sample:2000/01-2003/12.\\
In-Sample: 1985/01-1999.12. Short selling allowed.}
\begin{tabular}{lccc}
\toprule
Method & GMV-Turnover & MSR-Turnover & MV-Turnover \\
\midrule
NW & {\it 0.1136} & 0.1348 & 0.1624 \\
Residual-based NW & 0.3946 & 1.6363 & 0.4030 \\
POET & 0.3087 & 1.2184 & 0.3171 \\
OFT & 0.7746 & 17.4775 & 0.7580 \\
LSLW & 0.8796 & 3.4969 & 0.8796 \\
NLS & 0.4830 & 2.2657 & 0.4848 \\
SFNL & 0.4939 & 5.2308 & 0.4851 \\
S\&P 500 & {\bf 0.1036} & {\bf 0.1036} & {\bf 0.1036} \\
\bottomrule
\end{tabular}
\begin{tablenotes}
\small
\item GMV=Global minimum variance portfolio, MV=Mean-variance portfolio with target returns = 1\%. MSR=Maximum Sharpe ratio portfolio.Boldface number denotes the winner and italic one denotes the runner up.
\end{tablenotes}
\end{threeparttable}
\end{table}

\begin{table}[H]
\begin{threeparttable}
\centering
\caption{Turnover: Out-Sample:2008/01-2010/12.\\
In-Sample: 1993/01-2007/12. Short selling allowed.}
\begin{tabular}{lccc}
\toprule
Method & GMV-Turnover & MSR-Turnover & MV-Turnover \\
\midrule
NW & {\it 0.1103} & 0.1398 & 0.1327 \\
Residual-based NW & 0.3023 & 0.7511 & 0.3081 \\
POET & 0.2390 & 0.6955 & 0.2461 \\
OFT & 0.6516 & 2.5136 & 0.6380 \\
LSLW & 0.6637 & 2.0224 & 0.6647 \\
NLS & 0.4329 & 1.3818 & 0.4348 \\
SFNL & 0.4333 & 1.6782 & 0.4227 \\
S\&P 500 & {\bf 0.1011} & {\bf 0.1011} & {\bf 0.1011} \\
\bottomrule
\end{tabular}
\begin{tablenotes}
\small
\item  GMV=Global minimum variance portfolio, MV=Mean-variance portfolio with target returns = 1\%. MSR=Maximum Sharpe ratio portfolio.Boldface number denotes the winner and italic one denotes the runner up.
\end{tablenotes}
\end{threeparttable}
\end{table}

\begin{table}[H]
\begin{threeparttable}
\centering
\caption{Turnover: Out-Sample: 2020/01-2023/12.\\
In-Sample: 2005/01-2019/12. Short selling allowed.}
\begin{tabular}{lccc}
\toprule
Method & GMV-Turnover & MSR-Turnover & MV-Turnover \\
\midrule
NW & {\it 0.1120} & 0.1247 & 0.1397 \\
Residual-based NW & 0.3695 & 0.8085 & 0.3767 \\
POET & 0.2493 & 0.6158 & 0.2777 \\
OFT & 0.8859 & 2.4399 & 0.8475 \\
LSLW & 0.8571 & 1.8064 & 0.8520 \\
NLS & 0.5381 & 1.2218 & 0.5039 \\
SFNL & 0.5381 & 1.7272 & 0.5359 \\
S\&P 500 & {\bf 0.0823} & {\bf 0.0823} & {\bf 0.0823} \\
\bottomrule
\end{tabular}
\begin{tablenotes}
\small
\item GMV=Global minimum variance portfolio, MV=Mean-variance portfolio with target returns = 1\%. MSR=Maximum Sharpe ratio portfolio.Boldface number denotes the winner and italic one denotes the runner up.
\end{tablenotes}
\end{threeparttable}
\end{table}

\begin{table}[H]
\begin{threeparttable}
\centering
\caption{Turnover: Out-Sample: 2000/01-2010/12.\\
In-Sample: 1985/01-1999/12. Short selling allowed.}
\begin{tabular}{lccc}
\toprule
Method & GMV-Turnover & MSR-Turnover & MV-Turnover \\
\midrule
NW & {\it 0.0996} & 0.1210 & 0.1374 \\
Residual-based NW & 0.3033 & 1.0318 & 0.3101 \\
POET & 0.2402 & 0.8496 & 0.2485 \\
OFT & 0.6349 & 7.7697 & 0.6207 \\
LSLW & 0.7012 & 2.3047 & 0.7093 \\
NLS & 0.4019 & 1.5138 & 0.4078 \\
SFNL & 0.4049 & 2.7497 & 0.3970 \\
S\&P 500 & {\bf 0.0891} & {\bf 0.0891} & {\bf 0.0891} \\
\bottomrule
\end{tabular}
\begin{tablenotes}
\small
\item GMV=Global minimum variance portfolio, MV=Mean-variance portfolio with target returns = 1\%. MSR=Maximum Sharpe ratio portfolio.Boldface number denotes the winner and italic one denotes the runner up.
\end{tablenotes}
\end{threeparttable}
\end{table}

\begin{table}[H]
\begin{threeparttable}
\centering
\caption{Turnover: Out-Sample: 2000/01-2019/12.\\
In-Sample: 1985/01-1999/12. Short selling allowed.}
\begin{tabular}{lccc}
\toprule
Method & GMV-Turnover & MSR-Turnover & MV-Turnover \\
\midrule
NW & {\it 0.0965} & 0.1171 & 0.1260 \\
Residual-based NW & 0.2670 & 0.7848 & 0.2770 \\
POET & 0.1878 & 0.6242 & 0.1965 \\
OFT & 0.6356 & 5.1116 & 0.6196 \\
LSLW & 0.6766 & 1.3070 & 0.6826 \\
NLS & 0.3948 & 2.1386 & 0.3996 \\
SFNL & 0.4111 & 1.3801 & 0.3991 \\
S\&P 500 & {\bf 0.0778} & {\bf 0.0778}  & {\bf 0.0778} \\
\bottomrule
\end{tabular}
\begin{tablenotes}
\small
\item  GMV=Global minimum variance portfolio, MV=Mean-variance portfolio with target returns = 1\%. MSR=Maximum Sharpe ratio portfolio.Boldface number denotes the winner and italic one denotes the runner up.
\end{tablenotes}
\end{threeparttable}
\end{table}

\newpage

\section{Conclusions}

In this paper we consider seven methods that are used in ML-AI portfolio investment. Our main insight is that they have to be used jointly with three objective functions: Global Minimum Variance, Maximum Sharpe ratio, and Markowitz-Mean Variance. By analyzing three downturns, and two long time periods we see that Nodewise regression of \cite{caner2019} produces better results generally than other ML-AI and can beat $S\&P500$. The possible reason is that it is a model estimation free approach with consistency in high dimensions, and its structure can understand the conditional correlation between asset classes better.


\newpage

\bibliographystyle{chicagoa}
\bibliography{ai-invest}
\end{document}